\documentclass[]{ss_areviews}

\usepackage{ARAstroBib}

\begin{document}

\bibliographystyle{Astronomy}

\input epsf.tex    
\input epsf.def   
\input psfig.sty

\jname{Annual Reviews of Astronomy and Astrophysics}
\jyear{2009}
\jvol{47}
\ARinfo{481:522}


\newcommand\hi{H\,{\sc i}}
\newcommand\hei{He\,{\sc i}}
\newcommand\heii{He\,{\sc ii}}
\newcommand\lii{Li\,{\sc i}}
\newcommand\liii{Li\,{\sc ii}}
\newcommand\bei{Be\,{\sc i}}
\newcommand\beii{Be\,{\sc ii}}
\newcommand\bi{B\,{\sc i}}
\newcommand\ci{C\,{\sc i}}
\newcommand\niti{N\,{\sc i}}
\newcommand\oi{O\,{\sc i}}
\newcommand\nai{Na\,{\sc i}}
\newcommand\naii{Na\,{\sc ii}}
\newcommand\mgi{Mg\,{\sc i}}
\newcommand\mgii{Mg\,{\sc ii}}
\newcommand\ali{Al\,{\sc i}}
\newcommand\sii{Si\,{\sc i}}
\newcommand\siii{Si\,{\sc ii}}
\newcommand\phoi{P\,{\sc i}}
\newcommand\si{S\,{\sc i}}
\newcommand\ki{K\,{\sc i}}
\newcommand\cai{Ca\,{\sc i}}
\newcommand\caii{Ca\,{\sc ii}}
\newcommand\sci{Sc\,{\sc i}}
\newcommand\scii{Sc\,{\sc ii}}
\newcommand\tii{Ti\,{\sc i}}
\newcommand\tiii{Ti\,{\sc ii}}
\newcommand\vi{V\,{\sc i}}
\newcommand\vii{V\,{\sc ii}}
\newcommand\cri{Cr\,{\sc i}}
\newcommand\crii{Cr\,{\sc ii}}
\newcommand\mni{Mn\,{\sc i}}
\newcommand\fei{Fe\,{\sc i}}
\newcommand\feii{Fe\,{\sc ii}}
\newcommand\coi{Co\,{\sc i}}
\newcommand\coii{Co\,{\sc ii}}
\newcommand\nii{Ni\,{\sc i}}
\newcommand\cui{Cu\,{\sc i}}
\newcommand\zni{Zn\,{\sc i}}
\newcommand\srii{Sr\,{\sc ii}}
\newcommand\baii{Ba\,{\sc ii}}
\newcommand\euii{Eu\,{\sc ii}}

\newcommand{\loge}{$\log \epsilon$}
\newcommand{\deltaloge}{$\Delta \log \epsilon$}
\newcommand{\logli}{$\log \epsilon_{\rm Li}$}
\newcommand{\oh}{[O/H]}
\newcommand{\cfe}{[C/Fe]}
\newcommand{\nfe}{[N/Fe]}
\newcommand{\ofe}{[O/Fe]}
\newcommand{\co}{[C/O]}
\newcommand{\no}{[N/O]}
\newcommand{\alfe}{[Al/Fe]}
\newcommand{\crfe}{[Cr/Fe]}
\newcommand{\znfe}{[Zn/Fe]}
\newcommand{\bafe}{[Ba/Fe]}
\newcommand{\eufe}{[Eu/Fe]}

\newcommand\ao{{\it Applied Optics}}
\newcommand\aap{{\it Astron. Astrophys.}}
\newcommand\aapr{{\it Astron. Astrophys. Review}}
\newcommand\aaps{{\it Astron. Astrophys. Supp.}}
\newcommand\aj{{\it Astron. J.}}
\newcommand\an{{\it Astron. Nach.}}
\newcommand\apj{{\it Ap. J.}}
\newcommand\apjl{{\it Ap. J. L.}}
\newcommand\aplett{{\it Astroph. Letters.}}
\newcommand\apjs{{\it Ap. J. Supp.}}
\newcommand\araa{{\it Annu. Rev. Astron. Astrophys.}}
\newcommand\astrep{{\it Astron. Rep.}}
\newcommand\cjaa{{\it Chin. J. Astron. Astrophys.}}
\newcommand{\gca}{Geochim. Cosmochim. Acta}
\newcommand\jphysb{{\it J. Phys. B}}
\newcommand\jqsrt{{\it J.Q.S.R.T.}}
\newcommand\lr{{\it Living Reviews in Solar Physics}}
\newcommand\mnras{{\it MNRAS}}
\newcommand\memsait{{\it Mem.S.A.It.}}
\newcommand\nature{{\it Nature}}
\newcommand\newastrev{{\it New Astronomy Rev.}}
\newcommand\nat{{\it Nature}}
\newcommand\pasj{{\it P. A.S.J.}}
\newcommand\pasp{{\it P.A.S.P.}}
\newcommand\physscr{{\it Physica Scripta}}
\newcommand\physrep{{\it Physics Reports}}
\newcommand\pra{{\it Phys. Rev. A}}
\newcommand\physsoc{{\it Proc. Phys. Soc.}}
\newcommand\revmodphys{{\it Rev. Mod. Phys.}}
\newcommand\solphys{{\it Solar Physics}}
\newcommand\science{{\it Science}}
\newcommand\zphys{{\it Z. Phys.}}

\newcommand{\teff}{$T_{\rm eff}$}
\newcommand{\logg}{log\,$g$}
\newcommand{\feh}{[Fe/H]}
\newcommand{\micro}{$\xi_{\rm turb}$}
\newcommand{\macro}{$\zeta_{\rm macro}$}
\newcommand{\vsini}{$v_{\rm rot} \sin i$}
\newcommand{\vrot}{\mbox{$v_{\rm rot}$}}
\newcommand{\vrad}{$v_{\rm rad}$}
\newcommand{\kms}{km\,s$^{-1}$}
\newcommand{\sh}{$S_{\rm H}$}
\newcommand{\chiexc}{$\chi_{\rm exc}$}

\title{The chemical composition of the Sun}

\markboth{Asplund et al.}{The chemical composition of the Sun}

\author{
Martin Asplund 
\affiliation{
Max-Planck-Institut f\"ur Astrophysik, 
D-85741 Garching, Germany;
e-mail: asplund@mpa-garching.mpg.de
}
Nicolas Grevesse 
\affiliation{
Centre Spatial de Li\`ege, Universit\'e de Li\`ege,
B-4031 Angleur-Li\`ege, Belgium\\
Institut d'Astrophysique et de G\'eophysique, Universit\'e de Li\`ege,
B-4000 Li\`ege, Belgium;
e-mail: Nicolas.Grevesse@ulg.ac.be}
A. Jacques Sauval
\affiliation{Observatoire Royal de Belgique, 
B-1180 Bruxelles, Belgium;
e-mail: jacques.sauval@oma.be}
Pat Scott
\affiliation{Department of Physics and Oskar Klein Centre for Cosmoparticle Physics, Stockholm University, AlbaNova University Centre, 
S-10691 Stockholm, Sweden; e-mail: pat@fysik.su.se}
}


\begin{abstract}
The solar chemical composition is an important ingredient in our understanding
of the formation, structure and evolution of both the Sun and our solar system. 
Furthermore, it is an essential reference standard against which the elemental
contents of other astronomical objects are compared.
In this review we evaluate the current understanding of the solar photospheric composition.
In particular, we present a re-determination of the abundances of nearly all available elements, 
using a realistic new 3-dimensional (3D), time-dependent hydrodynamical model of the solar atmosphere.  
We have carefully considered the atomic input data and selection of spectral lines, and accounted 
for departures from LTE whenever possible. 
The end result is a comprehensive and homogeneous compilation of the solar elemental abundances.
Particularly noteworthy findings are significantly lower abundances of carbon, nitrogen, oxygen and neon 
compared with the widely-used values of a decade ago.
The new solar chemical composition is supported by a high degree of internal consistency 
between available abundance indicators, and by agreement with values obtained
in the solar neighborhood and from the most pristine meteorites. 
There is, however, a stark conflict with standard models of the solar interior according to helioseismology,
a discrepancy that has yet to find a satisfactory resolution.

\end{abstract}

\maketitle

\section{INTRODUCTION}
\label{s:intro}


The solar chemical composition is a fundamental yardstick in astronomy, to which
the elemental abundances of essentially all cosmic objects, be they planets, stars, nebulae
or galaxies, are anchored. The importance of having accurate solar elemental abundances 
can thus not be overstated. 
From the pioneering efforts of 
\citet{1929ApJ....70...11R}, 
\citet{1956RvMP...28...53S} 
and
\citet{1960ApJS....5....1G} 
to the more recent works of 
\citet{1989GeCoA..53..197A}, 
\citet{1998SSRv...85..161G}, 
\citet{2003ApJ...591.1220L}, 
\citet{2005ASPC..336...25A}, 
\citet{2007SSRv..130..105G} 
and
\citet{2009arXiv0901.1149L} 
compilations of the solar system abundances have found extremely wide-ranging use in astronomy and cosmology.

There are two independent and complementary ways of determining the solar system
abundances, each with its pros and cons. Through mass spectroscopy of meteorites in 
terrestrial laboratories, 
it is possible to directly measure the abundance of almost every element and isotope with
remarkable precision. The most pristine meteorites are the so-called CI chondrites,
which have been modified least by various physical and chemical processes over the past 4.56\,Gyr.
Even in these meteorites  volatile elements have been depleted to various degrees, 
including the six most abundant elements:
hydrogen, helium, carbon, nitrogen, oxygen and neon. As a consequence, it is
not possible to rely on meteorites to determine the primordial solar system abundances
for such elements. This also implies that one must measure all meteoritic abundances
relative to an element other than hydrogen, traditionally chosen to be silicon. 
Prior knowledge of the solar photospheric Si abundance  is therefore required in
order to place the meteoritic abundances on the same absolute scale as the Sun
\citep{1956RvMP...28...53S}. 

With the exception of a general $\sim 10$\%  modification due to diffusion and gravitational settling
(Sect. \ref{s:bulk})
and depletion of lithium and possibly beryllium, 
today's photospheric abundances are believed to reflect those at the birth of the solar system.
They are of profound importance for the understanding of our own solar system and
the Sun's interior structure and evolution.  
Unfortunately, photospheric abundances cannot be determined
with the same accuracy as those from meteorites. In particular, very little isotopic abundance
information is available.  Furthermore, photospheric abundances are not observed but
inferred. Thus the solar spectrum must be interpreted using realistic models
of the solar atmosphere and spectrum formation process.
In addition, the required atomic and molecular data
are rarely precise enough to rival the accuracy of
meteoritic measurements. 
In spite of the remaining uncertainties in solar photospheric 
determinations, having access to the most accurate solar abundances possible is crucial, 
not only for understanding our own Sun but also 
in order to make meaningful comparisons with other stars (for which 
meteoritic information is obviously unavailable). The Sun is also the only possible
reference for the interstellar medium, H\,{\sc ii} regions and other galaxies when
studying the elements depleted in meteorites.

In the decade since the appearance of the widely-used compilation of the solar
chemical composition by 
\citet{1998SSRv...85..161G}, 
the abundances of a large number of elements have been revised, several rather severely 
\citep[e.g.][]{2005ASPC..336...25A}. 
This is true in particular for C, N, O and Ne.
There are three main reasons for the changes:
improvements in atomic and molecular transition probabilities, the advent of
3D hydrodynamical solar model atmospheres and relaxation of
the assumption of local thermodynamic equilibrium in the spectral line formation. 
Several of the proposed abundance alterations have not gone unchallenged, however. 
It is therefore timely to now critically assess the reference solar
chemical composition and make recommendations regarding the most accurate values
available to the astronomical community.
After recapitulating the necessary ingredients in any solar abundance analysis
(Sect. \ref{s:ingredients}), we discuss in detail the solar chemical composition on 
an element-by-element basis (Sect. \ref{s:solarabund}). 
We also present a comparison between photospheric solar abundances and those inferred
by alternative means, such as via meteorites and stars or gas in the solar neighborhood 
(Sect. \ref{s:comparison}).
We conclude by summarizing the main points and outlining a few open issues
for the future (Sect. \ref{s:conclusions}).

\section{INGREDIENTS FOR SOLAR ABUNDANCE ANALYSIS}
\label{s:ingredients}

\subsection{Observations}
\label{s:observations}


Analyses of the solar photospheric chemical composition make use of the elemental
fingerprints present in the solar spectrum in the form of spectral absorption lines. 
There are several solar atlases of both intensity and flux spectra in the ultraviolet, 
optical and infrared regions. 
The most commonly used optical disk-center intensity spectra of the quiet Sun are the
so-called Jungfraujoch or Li\`ege atlas (after its locations of observation and production,
respectively; \citealt{1973apds.book.....D}) 
and the Kitt Peak solar atlas 
\citep{1984SoPh...90..205N}. 
The spectral resolving power of the latter is slightly higher while the former is less
affected by telluric absorption due to the higher observing altitude. 
There are also corresponding infrared disk-center intensity atlases observed from Kitt Peak
\citep{delbouille_ir} 
and from the {\sc atmos} experiment flown on the space shuttle
\citep{farmer_atmos, 1996ApOpt..35.2747A}. 
Finally, the solar flux spectrum has recently been re-reduced by
\citet{2006astro.ph..5029K}; 
see also \citet{1984sfat.book.....K} 
for an earlier rendition of the same Kitt Peak solar data set.
All solar atlases agree very well with each other except for spectral regions
afflicted by telluric features. Thus the quality of observations is in general
not a source of significant error in solar abundance analyses.

\subsection{Atomic and molecular data}
\label{s:inputdata}


A critical ingredient in any solar abundance analysis is of course the atomic and molecular
line input data. Of these the transition probability is the most obvious, but there are a raft of other
required components that are not always as well-determined as one would hope. 
These include line broadening, hyperfine and isotopic splitting, 
dissociation energies and partition functions. 
Much laudable work has been done in improving the $gf$-values for a large number of transitions
\citep[e.g.][]{
1992RMxAA..23...45K, 
2003ApJ...584L.107J, 
2006ApJS..162..227L, 
2007ApJ...667.1267S} 
but still several elements and lines have surprisingly uncertain transition probabilities.
Another important advancement in recent years is that it is now possible to accurately compute
the self-broadening (also known as van der Waals broadening) of metal lines 
\citep{
1995MNRAS.276..859A, 
2000A&AS..142..467B, 
2005A&A...435..373B}, 
which has made the classical Uns\"old broadening and 
its ad-hoc enhancement factors largely obsolete, at least for neutral species. 

The need for atomic data increases dramatically when relaxing the LTE assumption. 
In non-LTE calculations one needs transition data not only for the line in question but
in principle for all other lines of the element. In addition, photo-ionization cross-sections 
play a crucial role. There have been noteworthy improvements in this respect in recent years 
through the Opacity and Iron Projects 
\citep{2005MNRAS.360..458B} for example, 
but unfortunately the lack of quantum mechanical calculations for most elements preclude
detailed non-LTE investigations.
The most unsatisfactory situation, however, relates to collisional cross-sections for excitation and 
ionization, for which laboratory or theoretical calculations are completely missing
bar for a few elements and transitions relevant for spectroscopy of late-type stars
\citep[e.g.][]{2003PhRvA..68f2703B, 2007A&A...462..781B}. 
In the absence of realistic evaluations, classical recipes such as those of
\citet{1962ApJ...136..906V} 
and 
\citet{1968ZPhy..211..404D} 
are frequently used but cannot be expected to be more than order-of-magnitude estimates at best. 
Often scaling factors to these formulae are used, sometimes calibrated by achieving an
optimal overall fit to the various spectral features. 
It should remembered that such a procedure may reveal less about the
true collisional cross-sections than about other shortcomings in the modelling
of the stellar atmospheres and line formation
\citep{2005ARA&A..43..481A}. 

\subsection{Solar atmospheric and line formation modelling}
\label{s:models}


Extracting information about the Sun's chemical composition from the solar spectrum
requires realistic models of the solar atmosphere and the line formation process.
Traditional solar and stellar abundance analyses employ one-dimensional (1D) model
atmospheres assuming time-independence and hydrostatic equilibrium. 
Such models come in two flavours: theoretical and semi-empirical. The former are
constructed assuming a constant total (radiative plus convective) energy flux through
the atmosphere, which dictates the temperature stratification given the adopted 
equation-of-state and opacities. Convective energy transport is normally described
by the mixing length theory 
\citep{1958ZA.....46..108B}. 
The radiative flux is determined by solving the radiative transfer equation,
typically under the simplification of local thermodynamic equilibrium (LTE).
To achieve a realistic temperature structure, treatment of the combined effects 
of spectral lines (`line blanketing') must be as complete as possible.
The most widely-used theoretical 1D model atmospheres for solar-type stars are the 
\citet{1993ASPC...44...87K} 
and {\sc marcs} 
\citep{2008A&A...486..951G} 
grids of models.

\begin{figure}
\centerline{\psfig{figure=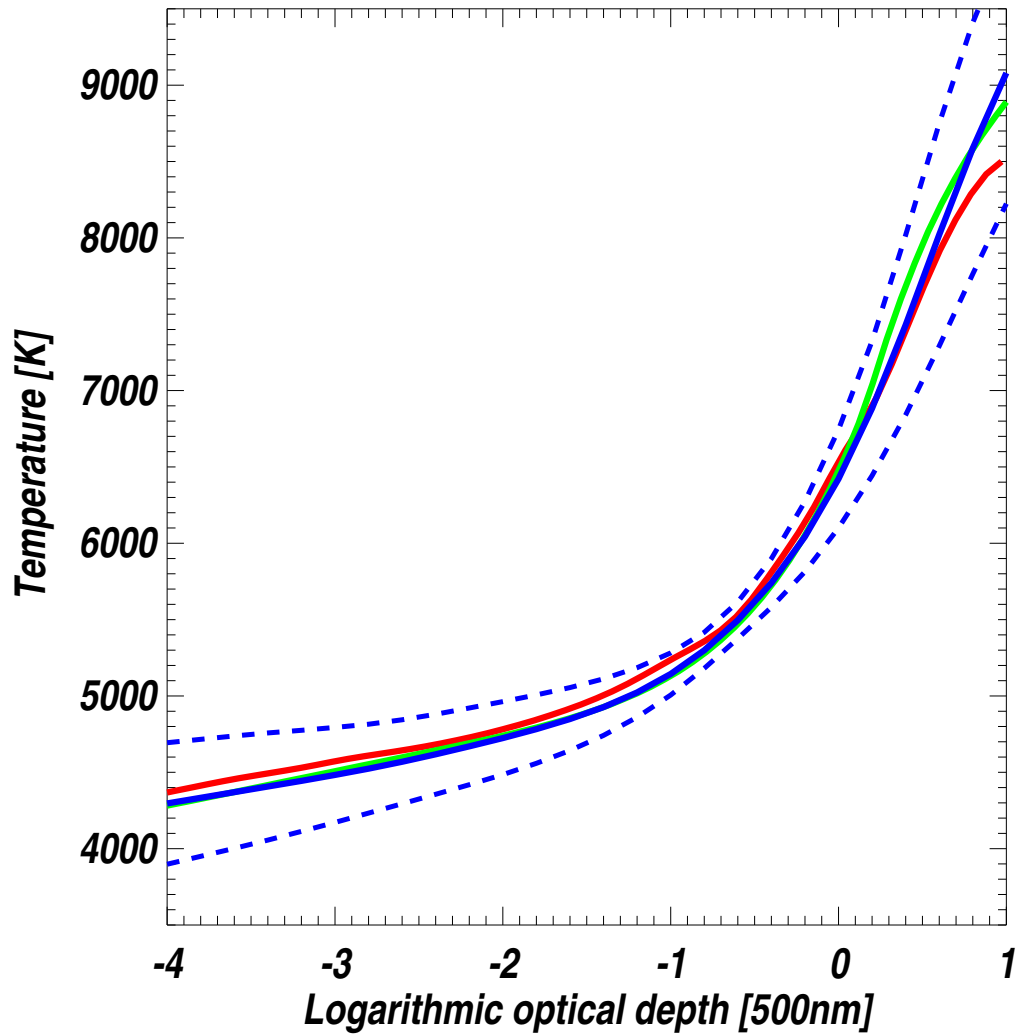,width=15cm}}
\caption{The mean temperature structure of the 3D hydrodynamical model of 
\citet{trampedach_sun}  
is shown as a function of optical depth at 500\,nm (blue solid line). The blue dashed
lines correspond to the spatial and temporal 
rms variations of the 3D model, while the red and green curves
denote the 1D semi-empirical
\citet{1974SoPh...39...19H} 
and the 1D theoretical {\sc marcs}
\citep{2008A&A...486..951G} 
model atmospheres, respectively.
\label{f:models}
}
\end{figure}

Semi-empirical models are often the preferred choice for solar spectroscopists.  In these models,
temperatures are inferred from depth-dependent observations such as the center-to-limb variation of the continuum
and/or lines with differing formation heights. 
The resulting atmospheric structure is thus sensitive to the ingredients involved in the
inversion process, such as LTE for the continuum and line formation, and the adopted line data.
Hydrostatic equilibrium is still assumed in semi-empirical models, but the flux constancy
criterion does not need to be obeyed nor is it necessary to estimate convective energy transport.
Several semi-empirical solar models are on the market, including the 
\citet{1974SoPh...39...19H}, 
{\sc val3c}
\citep{1976ApJS...30....1V} 
and {\sc miss}
\citep{2001ApJ...558..830A} 
models. The favoured model of solar abundance aficionados is the
\citeauthor{1974SoPh...39...19H} 
model, partially because its temperature stratification is expected to be an accurate
representation of the real photosphere -- but also partially just by habit.
Indeed it is somewhat surprising that the {\sc val3c} model 
or its successors 
\citep[e.g.][]{2006ApJ...639..441F} 
have not caught on more within the solar abundance community, since they are built on 
similar premises but without the LTE restriction. 
The venerable 
\citeauthor{1974SoPh...39...19H} 
model in fact dates back more than forty years, 
since the 1974 version was simply a new pressure-integration of the original 
\citet{1967ZA.....65..365H}  
model using updated equation-of-state and continuous opacities. 
It was constructed to match the observed continuum center-to-limb variation 
as well as the line depths of some 900 lines of varying strengths, 
under the assumption of LTE. Due to the limited spectral resolution of the solar spectrum used by
\citeauthor{1967ZA.....65..365H}, 
one would expect the temperatures to be overestimated by about 50\,K in the line-forming region.
Interestingly, the corrected version of such a temperature structure 
would be more similar to the {\sc miss} semi-empirical model 
\citep{2001ApJ...558..830A}; 
both semi-empirical models have a shallower temperature gradient than 
1D theoretical model atmospheres, as seen in Fig. \ref{f:models}.
It should be borne in mind that there exists an ambiguity when employing any 1D model: 
while many simply adopt temperatures, pressures and densities directly from the published model,
a more correct procedure is to perform a pressure-integration using only the temperature structure as a 
function of continuum optical depth, 
in order to be consistent with the opacities computed with the spectrum synthesis code one employs.

More recently it has become tractable to perform 3D, time-dependent, hydrodynamical simulations
of the stellar surface convection
\citep[e.g.][]{
1998ApJ...499..914S, 
1999A&A...346L..17A, 
2002AN....323..213F, 
2004A&A...421..741V}. 
Here the standard conservation equations of
mass, momentum and energy are solved together with the 3D radiative transfer equation in a small
but representative volume of the stellar atmosphere. 
Radiative energy exchange with the gas
in the surface layers is crucial, because this is what drives the convective motions. Because of
the much more computationally demanding nature of such hydrodynamical modelling,
approximations must be made in the radiative transfer treatment. Rather than solving
for many thousands of frequency points as regularly done in theoretical 1D hydrostatic models, 
the wavelength variation of opacity is represented by a small number of typical opacity bins (4-20) in order
to compute the radiative heating/cooling rate that enters the energy equation
\citep{1982A&A...107....1N}.
Detailed comparisons with the monochromatic solution reveal that this scheme is surprisingly accurate.
The reader is referred to 
\citet{2009LRSP....6....2N} 
for a recent review of the physics of solar surface convection and the numerical details of its simulation. 

A few different 3D hydrodynamical solar model atmospheres 
have been constructed and employed for abundance purposes
\citep[e.g.][]{
2000A&A...359..729A, 
2008A&A...488.1031C, 
trampedach_sun}. 
While these are computed with different codes, they are based on many of the same physical assumptions and
approximations.  It is therefore not too surprising that the resulting atmospheric structures are similar. 
The 3D models have been computed using realistic and comprehensive equation-of-state, 
continuous and line opacities. 
The numerical resolution is sufficiently high to adequately resolve the granulation, 
albeit obviously not high enough to reach down to the microscopic viscous dissipation length-scales.
The reason that these so-called `large-eddy' simulations are nevertheless 
quite successful in reproducing the real photosphere
is that the mean stratification is largely independent of numerical resolution
\citep{1998ApJ...499..914S, 2000A&A...359..669A}. 
Spectral line formation is
heavily biased to the upflows, which have a low degree of turbulence in 
comparison to the downflows, due to the divergent flow. 
Obviously none of the mixing length parameters required in 1D models enter into 3D simulations, as the
convective energy flux is instead a natural consequence of the hydrodynamics. 

Having a suitably realistic solar model atmosphere, the next step when trying to infer 
the solar chemical composition is to model how the spectrum is formed. 
In order to derive an abundance of an element, solar spectroscopists either
measure the total observed line strength as defined by the equivalent width, 
or directly compare the observed and theoretical spectra through the fitting of line profiles. 
Finally, to compute the line strength requires knowledge of the level populations. 
In LTE, these follow straightforwardly from the local temperature and electron pressure
using the Boltzmann and Saha distributions. 
In general this rather severe simplification cannot be expected to be valid, and the non-LTE
case must instead be considered. While non-LTE is a general term covering all situations when LTE is not valid,
in practice all calculations referred to herein assume statistical equilibrium, i.e. the level populations
do not change with time. 
Non-LTE thus requires a simultaneous solution of the rate equations for all relevant levels and species
together with the radiative transfer equation for all relevant wavelengths.
We refer to 
\citet{2005ARA&A..43..481A} 
for a detailed account of the principles of non-LTE and the most important non-LTE effects 
encountered for late-type stars like the Sun.

\subsection{Observational constraints on solar modelling}
\label{s:obstests}

Since more sophisticated modelling does not automatically imply a more realistic outcome,
it is of paramount importance to carefully test the predictions of the 3D model atmospheres against an arsenal of 
observational diagnostics. 
This is especially true given that 1D model atmospheres more often than not fail the very same tests. 
Current generations of 3D models are very successful in reproducing the observed
solar granulation topology, typical length- and time-scales, convective velocities and 
intensity brightness contrast
\citep[e.g.][]{1998ApJ...499..914S, 2009LRSP....6....2N}, 
clearly none of which 1D models are able to predict.

\begin{figure}
\centerline{\psfig{figure=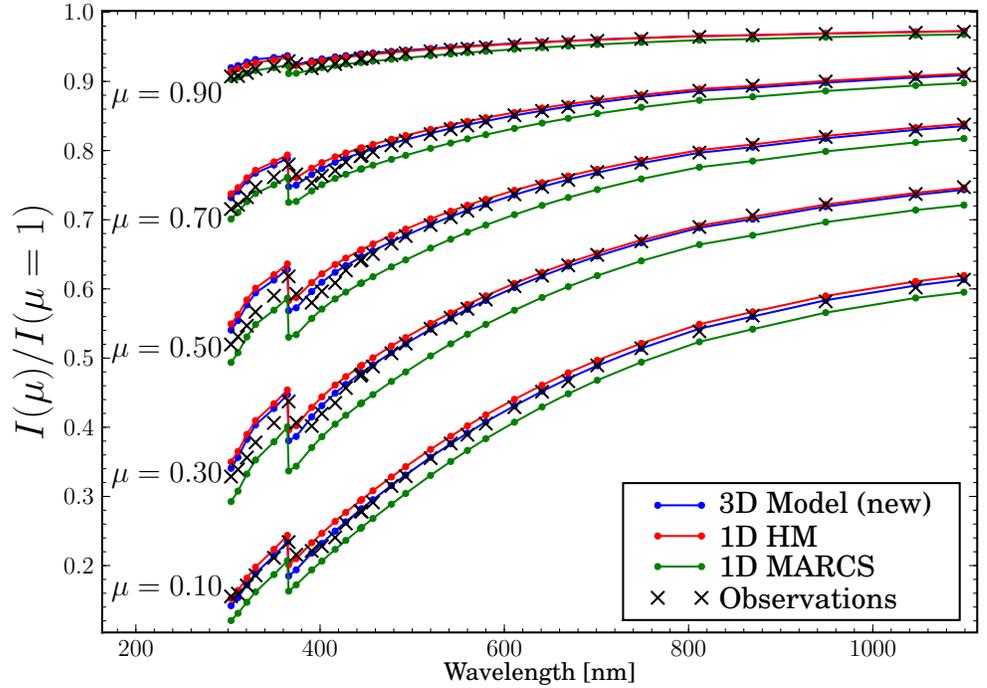,width=15cm}}
\caption{Comparison of the predicted continuum center-to-limb variation as a function
of wavelength for different solar model atmospheres against observations
\citep{1994SoPh..153...91N}. 
The results for five different viewing angles $\mu$
are shown from near disk center  ($\mu =0.9$) to close to the limb ($\mu =0.1$).
All intensities are normalized to the corresponding disk-center intensities.
The 3D hydrodynamical model 
\citep{trampedach_sun}  
outperforms even the 
\citet{1974SoPh...39...19H} 
semi-empirical model, which was designed to satisfy this diagnostic
\citep{pereira_models}. 
As for all 1D theoretical model atmospheres, the {\sc marcs} model has a too steep
temperature gradient, which manifests itself in a poor agreement with the center-to-limb variation.
\label{f:clv}
}
\end{figure}

Another important constraint is the center-to-limb variation of the continuum as a function of wavelength.
\citet{2006ApJS..165..618A} 
have criticized the 3D hydrodynamical model of 
\citet{2000A&A...359..729A}, 
oft used in recent solar abundance analyses, for having 
too steep a temperature gradient as judged by the observed center-to-limb variation.
\citet{2008ApJ...680..764K} 
reach a somewhat milder conclusion, presumably because they performed spectral 
synthesis with the full 3D model rather than with only the spatially-averaged version used by
\citeauthor{2006ApJS..165..618A}. 
The 3D {\sc co5bold} solar model of 
\citet{2008A&A...488.1031C} 
appears to perform better in this respect (H. Ludwig, private communication). 
In the present work, we make extensive use of the 3D solar model by
\citet{trampedach_sun},  
which has been constructed with an improved treatment of the radiative transfer
and updated opacities. 
Because of its somewhat shallower temperature stratification than the 3D model of 
\citeauthor{2000A&A...359..729A} 
(Fig. \ref{f:models}), 
the new 3D model satisfies the center-to-limb variation constraint very well,
as illustrated in Fig. \ref{f:clv}.
Indeed it outperforms all 1D models, even the semi-empirical
\citep{1974SoPh...39...19H} 
model, which was designed to fit this diagnostic
\citep{pereira_models}. 

The wings of the hydrogen Balmer lines are also sensitive tracers of the
temperature stratification in the deeper layers of the photosphere.
In recent years, significant improvements have been made in 
the treatment of the broadening of H lines, especially self-broadening
\citep{2000A&A...363.1091B, 2008A&A...480..581A}. 
It is still unclear whether noticeable non-LTE effects can be expected in the solar
atmosphere, since available calculations for the cross-sections of inelastic H$+$H collisions
are uncertain
\citep{2007A&A...466..327B}. 
With 1D theoretical model atmospheres, the H lines imply too low \teff\ for the Sun by
$50-100$\,K, and have the wrong line shapes
when the most accurate line broadening data are used
\citep{2002A&A...385..951B}. 
Such problems can be partly cured by playing with the mixing length parameters, however. 
In contrast, the H lines suggest that the temperature gradient of the 
\citet{1974SoPh...39...19H} 
semi-empirical model is too shallow.
\citet{pereira_models} 
found that the 3D solar model of 
\citet{trampedach_sun}  
reproduces the H$\alpha$ and H$\beta$ lines very well in LTE, 
doing so significantly better than the 1D models they investigated, but without requiring the tweaking of any
free parameters.
The {\sc co5bold} model has to our knowledge not yet been tested against H lines.

\begin{figure}
\centerline{\psfig{figure=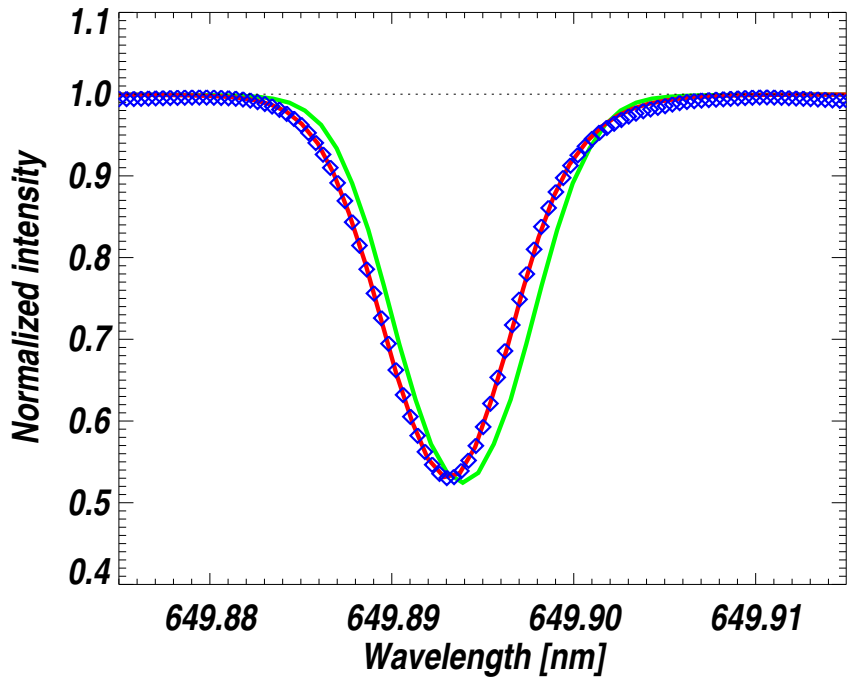,width=13cm}}
\caption{The predicted spectral line profile of a typical \fei\ line from 
the 3D hydrodynamical solar model (red solid line) compared with the observations (blue rhombs).
The agreement is clearly very satisfactory, which is the result of the Doppler shifts arising 
from the self-consistently computed convective motions that
broaden, shift and skew the theoretical profile. For comparison purposes also 
the predicted profile from a 1D model atmosphere (here
\citealt{1974SoPh...39...19H}) 
is shown; the 1D profile has been computed with a microturbulence of 1\,\kms\ and a
tuned macroturbulence to obtain the right overall linewidth. Note that even with these two free
parameters the 1D profile can neither predict the shift nor the asymmetry of the line. 
\label{f:lineprofiles}
}
\end{figure}

As seen in Fig. \ref{f:lineprofiles}, with 3D hydrodynamical models for the solar atmosphere it
is now possible to achieve highly satisfactory agreement with observed profiles of typical
weak and intermediate strong lines (the wings of strong lines reveal more about the
pressure broadening data than the employed model atmosphere). 
It is important to remember that this is achieved without invoking
the free parameters necessary in any 1D analysis (micro- and macroturbulence) in the line formation
calculations.
Such line broadening can thus be explained as the result of the Doppler shift arising from
the convective motions with a smaller contribution from the solar oscillations
\citep[e.g][ and references therein]{2009LRSP....6....2N}. 
Indeed, even the observed line shifts and asymmetries are very well reproduced with 3D models
\citep[e.g.][]{2000A&A...359..729A}. 
The {\sc co5bold} solar model produces good overall line profiles 
\citep[e.g.][]{2008A&A...488.1031C} 
but whether this extends also to line asymmetries has not yet been investigated.

In summary, the 3D hydrodynamical model atmosphere 
\citep{trampedach_sun}  
that we draw results from in this review outperforms 
1D model atmospheres in the available observational tests, from granulation topology and center-to-limb variation
to H lines and the detailed profiles of metallic lines.
For the purposes of photospheric abundance determinations, 
the 3D solar model employed here appears to be a very realistic representation of the solar photosphere.
No large differences are expected from using our 3D model or the corresponding {\sc co5bold} model 
\citep{2008A&A...488.1031C} 
given their overall similarities.


\section{PHOTOSPHERIC ABUNDANCES}
\label{s:solarabund}

\begin{figure}
\centerline{\psfig{figure=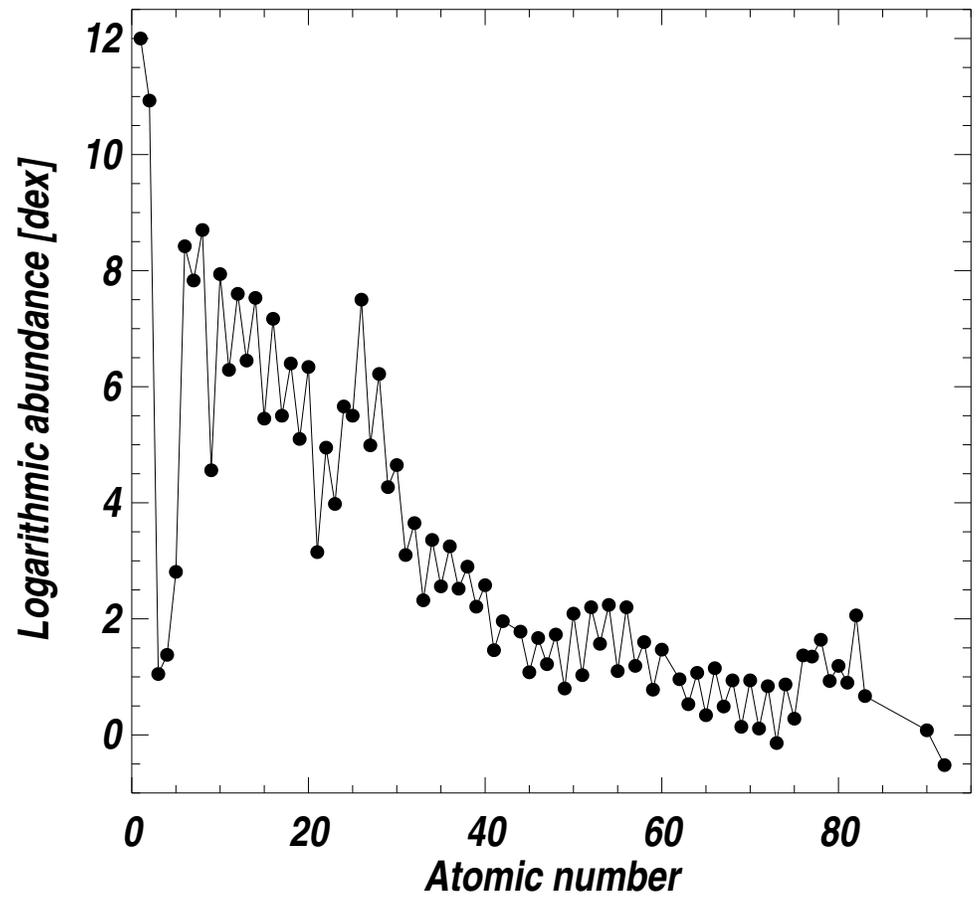,width=15cm}}
\caption{The present-day solar photospheric elemental abundances as a function of atomic number. 
As throughout this article, the logarithmic abundance of hydrogen is defined to be $\log \epsilon_{\rm H} = 12.0$.
\label{f:abund}
}
\end{figure}

Our recommended solar photospheric elemental abundances are listed in Table \ref{t:sun}, which
also provides the corresponding meteoritic values for CI chondrites to be discussed
in Sect. \ref{s:meteorites}.
[Throughout this review, we adopt the customary astronomical scale for logarithmic abundances
where hydrogen is defined to be $\log \epsilon_{\rm H} = 12.00$, i.e. 
$\log \epsilon_{\rm X} = \log (N_{\rm X}/N_{\rm H}) + 12$, where
$N_{\rm X}$ and $N_{\rm H}$ are the number densities of element X and hydrogen, respectively.]
Fig. \ref{f:abund} shows how the solar abundances vary with atomic number, illustrating several key 
features of nuclear and stellar physics: 
the high primordial H and He abundances; the fragile nature of Li, Be and B;
the relatively high abundance of elements involved in stellar H-, He- and C-burning,
modulated by the odd-even effect and the $\alpha$-capture effect;
the high nuclear binding energy and near nuclear statistical equilibrium for the Fe-peak;
and the production of the heavy elements through successive neutron capture
with peaks around the magic nuclei possessing closed neutron shells 
\citep[e.g.][]{1997nceg.book.....P}. 

In this review, we have attempted to reanalyse the solar abundances of
(nearly) all elements in a homogeneous manner with the best possible atomic data and
state-of-the-art solar modelling.
As a guiding principle, we have been very discerning when selecting the
lines for each element, since inclusion of dubious lines only increases the abundance scatter
and tends to skew the results towards higher abundances due to blends. 
The analysis has been carried out 
using several 1D and 3D model atmospheres and with non-LTE
effects accounted for whenever possible, all done with the same well-tested computer codes. 
Unless specified otherwise, the results presented here have been based on the
3D hydrodynamical solar model atmosphere of
\citet{trampedach_sun}. 
Below we present in some detail how the abundances were derived but the 
full description of the analysis (including line lists with all the relevant data for the transitions)
will appear in a forthcoming series of articles in {\em Astronomy \& Astrophysics}
\citep{sun_c, sun_n, sun_o, sun_na-ca, sun_fepeak, sun_heavy}. 

There are a wide range of potential
sources of error in solar and stellar abundance analyses, from inaccurate input data
for the transition ($gf$-values, line broadening etc) to difficulties in estimating line strengths 
(finite $S/N$, continuum placement, blends etc), to inadequacies in the atmospheric
and line-formation modelling (1D, LTE, mixing length and microturbulence parameters, continuous opacities, etc).
No consensus exists in the solar abundance literature on how the uncertainties should be quantified.
Some authors have used the standard deviation arising from the chosen sample of lines to 
represent the total error for a given species whilst others have employed the standard error of the mean or
some different measure altogether.
Previous compilations of
the solar chemical composition contained a mixture of different error estimates,
since the values were taken from sources with differing modus operandi.

We have attempted to quantify three possible systematic errors
introduced by the modelling: mean atmospheric stratification, atmospheric inhomogeneities and
departures from LTE. The first uncertainty has been estimated by taking half the difference
between the results with the spatially-averaged 3D model (denoted $<$3D$>$) and the 
\citet{1974SoPh...39...19H} 
model. The second has been evaluated as half the difference between 
the full 3D results and those from $<$3D$>$.
The inclusion of the theoretical {\sc marcs}
\citep{2008A&A...486..951G} 
and semi-empirical {\sc miss}
\citep{2001ApJ...558..830A} 
models serves as a further check on these two systematic errors.
Due to the general lack of non-LTE calculations for the majority of elements, the non-LTE
uncertainty is more difficult to quantify. Somewhat arbitrarily, in most cases we have  
chosen half of the predicted non-LTE abundance correction as an estimate of the error.
Occasionally we have employed the difference between
the cases with and without inelastic H collisions through the 
\citet{1968ZPhy..211..404D} 
formula, whilst taking into account the recommendations by
\citet{2005ARA&A..43..481A}. 
We have used an arbitrary, minimal estimate of the possible errors in the non-LTE corrections of 0.03\,dex.
In some cases this might overestimate the non-LTE systematic error, whilst in others (e.g.~\tii) it probably
underestimates the error; there is unfortunately no other consistent way to include such unknown errors. 
Estimating possible systematic errors arising from inappropriate line input data 
has not been attempted here; relative errors in $gf$-values
will at least be accounted for in the dispersion of the lines.
When the elemental abundance is determined by only one or very few lines, we have attempted
to estimate an additional error arising from uncertainties due to continuum placement and possible blends.
The systematic errors have been added in quadrature with the statistical error to
estimate the total uncertainty, which is used throughout the article unless specified otherwise.
The statistical error has been computed from the weighted standard error of the mean, with
weights assigned to each line according to uncertainties in the continuum placement and 
known or suspected minor blends. 

\subsection{Lithium, Beryllium and Boron}
\label{s:libeb}


{\bf Lithium:}
In the Sun, Li is depleted by about a factor of 150 compared with the already small meteoritic value. 
Even the \lii\ resonance line at 670.8\,nm is therefore very weak. The line
is blended by an assortment of CN and \fei\ lines, which makes the Li abundance determination
very challenging. 
Because the transition arises from the ground state of a minority species, the line is
sensitive to both the temperature and photo-ionizing UV radiation field
\citep{1997ApJ...489L.107K, 
2005ARA&A..43..481A}. 
The recommended value given in Table \ref{t:sun} stems from the analysis of 
\citet{1975SoPh...41...53M}, 
which has been corrected for the effects of atmospheric inhomogeneities and 3D non-LTE effects
\citep{
1999A&A...346L..17A, 
2003A&A...409L...1B}. 
The solar Li depletion cannot be explained by standard models based on the mixing length theory
of convection but requires additional mixing below the convection zone
\citep{1999ApJ...525.1032B, 2005Sci...309.2189C}. 

{\bf Beryllium:}
Early work implied that Be was depleted in the solar convection zone by about a factor of two
\citep{1975A&A....42...37C}. 
This interpretation was challenged by
\citet{1998Natur.392..791B}, 
who argued that one must take into account the long-suspected missing UV opacity 
\citep[e.g.][ and references therein]{2005ApJ...618..926S} 
in synthesizing the \beii\ resonance lines at 313\,nm. 
They estimated the amount of missing opacity by requiring that the nearby 
OH electronic lines would return the same O abundance as the OH vibrational lines in the infrared, which 
increased the derived photospheric Be abundance to the meteoritic value. 
The required extra opacity can be largely attributed to photo-ionization of \fei\ 
\citep{2001ApJ...546L..65B}. 
The
\citeauthor{1998Natur.392..791B} 
analysis was corroborated by 
\citet{2004A&A...417..769A}, 
who found that the basic conclusions are independent of the adopted model atmosphere, whether 1D or 3D.
The \beii\ lines are not  sensitive to non-LTE effects in the Sun
\citep{2005ARA&A..43..481A}. 

{\bf Boron:}
The problem of the missing UV opacity may also impact the solar B abundance 
determination, which can realistically only be based on the \bi\ 249.7\,nm resonance line. 
\citet{1999ApJ...512.1006C} 
have re-analysed the line with careful consideration of the important
\mgi\ photo-ionization cross-sections.
The effects of departures from LTE
\citep{1996A&A...311..680K} 
and 3D hydrodynamical model atmospheres
\citep{2005ASPC..336...25A} 
are small and opposite in sign. 
There is still a large uncertainty attached to the 
resulting abundance, $\log \epsilon_{\rm B} = 2.70 \pm 0.20$, 
due to the difficulty in analysing this crowded spectral region. 
Not surprisingly, B does not appear depleted in the solar photosphere. 

\subsection{Carbon}
\label{s:carbon}


The solar carbon abundance can be inferred from a wealth of indicators, from low-excitation forbidden and
high-excitation permitted atomic lines to various molecular transitions of CH, C$_2$ and CO
\citep{1978MNRAS.182..249L}. 
This fortunate circumstance has not translated into a well-determined abundance over 
the years as the different diagnostics have often yielded discrepant results.
The preliminary abundance $\log \epsilon_{\rm C} = 8.56 \pm 0.04$ recommended by
\citet{1989GeCoA..53..197A} 
eventually became $8.60 \pm 0.05$ in the final analysis of 
\citet{1991A&A...242..488G}. 
\citet{1998SSRv...85..161G} 
based their value of $8.52 \pm 0.06$ on an unpublished analysis employing a 
\citet{1974SoPh...39...19H} 
model with a modified temperature structure, in an attempt to remove existing trends with
excitation potential and equivalent widths for C, N, O and Fe.
Over the past decade the solar C abundance has undergone an even more drastic downward revision.
\citet{2002ApJ...573L.137A} 
reanalysed the [\ci ] 872.7\,nm line and found that the combination of 
a 3D hydrodynamical solar model 
\citep{2000A&A...359..729A} 
and an updated $gf$-value brought the abundance down to $8.39 \pm 0.04$.
The same value was obtained by 
\citet{2005A&A...431..693A}, 
who also employed \ci , C$_2$ and CH electronic and vibration lines
using the same 3D model. Convincingly, all indicators gave consistent results, further 
strengthening the case for the reality of the low C abundance.
In sharp contrast, the abundances estimated with the 
\citet{1974SoPh...39...19H} 
model atmosphere disagreed by $\approx 0.2$\,dex. 
The presence of temperature inhomogeneities and 
a cooler mean temperature structure in the 3D model compared with the previously employed
\citet{1974SoPh...39...19H} 
model reduced the deduced abundance from the molecular lines in particular, while significant
departures from LTE had a similar effect on the \ci\ lines. 
Using weak lines of the extremely temperature-sensitive CO molecule
and a self-consistently derived O abundance, 
\citet{2006A&A...456..675S} 
corroborated this finding.
They also determined an isotopic ratio of $^{12}{\rm C}/^{13}{\rm C} = 86.8 \pm 3.8$.

The value we recommend here, $\log \epsilon_{\rm C} = 8.43 \pm 0.05$, is slightly higher than in 
\citet{2005ASPC..336...25A} 
and stems from the mean of the 3D-based results for [C\,{\sc i}], C\,{\sc i}, CH and C$_2$ lines
\citep{sun_c}. 
As illustrated in Table \ref{t:cno}, the various C indicators imply highly uniform
abundances. With the exception of a correlation between
derived abundance and line strength for \ci, 
there are no trends with transition properties such as excitation potential 
or equivalent width for any of the species. 
A similar trend was also present in 
\citet{2005A&A...431..693A}, 
which was attributed to underestimated non-LTE effects for the stronger lines
\citep{2006A&A...458..899F}.
We note that only 1D non-LTE abundance corrections have been employed for
the C\,{\sc i} results while the
departures from LTE are expected to be exacerbated in the presence of atmospheric inhomogeneities
\citep{2005ARA&A..43..481A}; 
a full 3D non-LTE study for C would clearly be worthwhile. 
Given the importance of this abundant element, it is essential that the results of
\citet{sun_c} 
are confirmed by an independent group using an independent 3D model, akin to the 
analysis of the solar O abundance
\citep{2008A&A...488.1031C}. 

\subsection{Nitrogen}
\label{s:nitrogen}


The nitrogen abundance can be determined from
both atomic and molecular lines, although the forbidden [\niti ] lines are too weak in
the solar spectrum to be measurable
\citep{1990A&A...232..225G}. 
A preliminary N abundance was given in 
\citet{2005ASPC..336...25A} 
using high excitation \niti\ lines and NH vibration-rotation lines
in the infrared while in the final analysis
\citep{sun_n} 
also NH pure rotation lines and a multitude of transitions from various bands of CN were
considered in the framework of an improved 3D model.
The non-LTE effects for \niti\ are relatively small: $\approx -0.05$\,dex when ignoring H collisions
\citep{2009A&A...498..877C}. 
In their own analysis of these very weak \niti\ lines using their {\sc co5bold} 3D solar model, 
\citet{2009A&A...498..877C} 
obtained a \niti -based abundance which is 0.05\,dex higher than the corresponding value in 
\citet{sun_n}, 
largely due to the selection of lines. 
They included many lines that we consider to be of dubious value due to known or suspected blends. 
This is also reflected in the dramatically different standard deviations for the \niti\ lines in the two studies:
$\sigma = 0.04$ and $0.12$\,dex, respectively.
\citet{sun_n} 
found very good agreement between abundances inferred from 
\niti\ and NH lines, with very small dispersion in the NH results. 
The different CN bands also gave very similar results when employing a self-consistently derived
C abundance with the 3D model.  Unfortunately, 
\citet{2009A&A...498..877C} 
did not consider molecular lines and it is therefore not known how well their 
3D model performs in achieving consistent results for atoms and molecules.
It is sometimes stated in the literature that the molecular lines are more temperature sensitive
than atomic lines, but in fact the NH lines are less so than these highly excited \niti\ lines.

The N abundance from 
\citet{sun_n} 
that we recommend here, 
$\log \epsilon_{\rm N} = 7.83 \pm 0.05$,
is 0.09\,dex lower than suggested by 
\citet{1998SSRv...85..161G} 
and 0.22\,dex lower than in
\citet{1989GeCoA..53..197A}. 

\subsection{Oxygen}
\label{s:oxygen}


Oxygen is the most abundant element in the cosmos
not produced in the Big Bang, and the third most common overall.
Oxygen and its isotopes are key tracers of the formation and evolution
of planets, stars and galaxies and as such it is one of the most important
elements in all of astronomy.
Nevertheless, or perhaps consequentially,
its abundance often seems to be in dispute with 
the solar O abundance being no exception.
The favoured solar content has dropped almost precipitously over
the past two decades from 
$\log \epsilon_{\rm O} = 8.93 \pm 0.04$ in
\citet{1989GeCoA..53..197A} 
to 8.66 in
\citet{2005ASPC..336...25A}. 
The value we recommend here, $\log \epsilon_{\rm O} = 8.69 \pm 0.05$, is taken from 
\citet{sun_o}, 
who presented a 3D-based study of the available diagnostics:
forbidden [\oi ] and permitted \oi\ transitions as well as vibration-rotation and pure rotation lines of OH.

\citet{2001ApJ...556L..63A}  
reanalysed the [\oi ] 630\,nm line with a 3D model atmosphere. 
Based on the line profile they demonstrated 
that the line is blended by a \nii\ line, which
was subsequently confirmed experimentally
\citep{2003ApJ...584L.107J}. 
In combination with the 3D model, recognition of the blend led to a substantial
lowering of the solar O abundance compared with the canonical value at the time. 
\citet{2008ApJ...686..731A} 
revisited the 630\,nm line with a single snapshot of another 3D model but
found a 0.12\,dex higher O abundance, partly because he allowed the \nii\ $gf$-value
to vary freely in spite of the new accurate measurement by 
\citet{2003ApJ...584L.107J}. 
\citet{2008A&A...488.1031C} 
used several snapshots from the same {\sc co5bold} 3D model but instead 
fixed the Ni contribution assuming the Ni abundance from 
\citet{1998SSRv...85..161G} 
thereby confirming the results of 
\citet{2001ApJ...556L..63A}  
and 
\citet{2004A&A...417..751A}. 
A more self-consistent approach is to first determine the Ni abundance before
considering the [\oi ] line, as done by 
\citet{2009ApJ...691L.119S} 
with the same 3D model as employed by 
\citet{2004A&A...417..751A}. 
As explained by 
\citeauthor{2009ApJ...691L.119S}, 
it is then possible to very accurately predict the \nii\ contribution to the overall
line strength, independent of the reference atmospheric model.
\citet{sun_o} 
updated this analysis 
and found that the effects of the increased \nii\ proportion and their shallower model temperature gradient 
in comparison with 
\citet{2001ApJ...556L..63A}  
compensate for one another, leaving the abundance at 
$\log \epsilon_{\rm O} = 8.66$.
This abundance is also consistent with the center-to-limb variation of the 630\,nm feature
\citep{pereira_clv}. 

In addition to the 630\,nm line, there are two other [\oi ] transitions that can be utilized. 
The 636\,nm line is located in the midst of a very broad \cai\ auto-ionization line as well
as blended by CN lines; the latter's contribution can however be accurately predicted from
other neighboring CN lines. 
Both 
\citet{2008A&A...488.1031C} 
and
\citet{sun_o} 
found that this line returns a slightly higher abundance than the 630\,nm line, but
still consistent within the uncertainties. 
\citet{2008A&A...490..817M} 
added the [\oi ] 557\,nm line, whose slightly excited lower level makes it
largely independent of the model atmosphere. 
Unfortunately this transition is badly blended by C$_2$ lines, whose contribution 
\citeauthor{2008A&A...490..817M} 
attempted to estimate using nearby C$_2$ lines with the same line strengths and
excitation potential. 
\citet{sun_o} 
obtained an abundance from the 557\,nm line exactly 
in between those from the other two forbidden transitions.
The three [\oi ] lines imply a solar O abundance of 
$\log \epsilon_{\rm O} = 8.70 \pm 0.05$ when due consideration is given to the blends.

Recently,
\citet{2008ApJ...682L..61C} 
introduced a novel method to estimate the O abundance, namely
spectropolarimetry of the [\oi ] 630\,nm line in a sunspot. Because of the different
polarization of the [\oi] and \nii\ lines it is possible to disentangle the \nii\ contribution. 
After consideration of the fraction of O tied up in CO in the cool environments of sunspots, 
they obtained a high O abundance of $\log \epsilon_{\rm O} = 8.86$.
\citet{2009ApJ...691L.119S} 
have however pointed out that 
\citeauthor{2008ApJ...682L..61C} 
applied an outdated $gf$-value for the [\oi ] line.
Furthermore, they showed that when using an improved Ni abundance 
and an alternative CO treatment one finds a significantly lower O abundance:
$\log \epsilon_{\rm O} = 8.71$; a further reduction is obtained when employing an
alternative sunspot model. 

\begin{figure}
\centerline{\psfig{figure=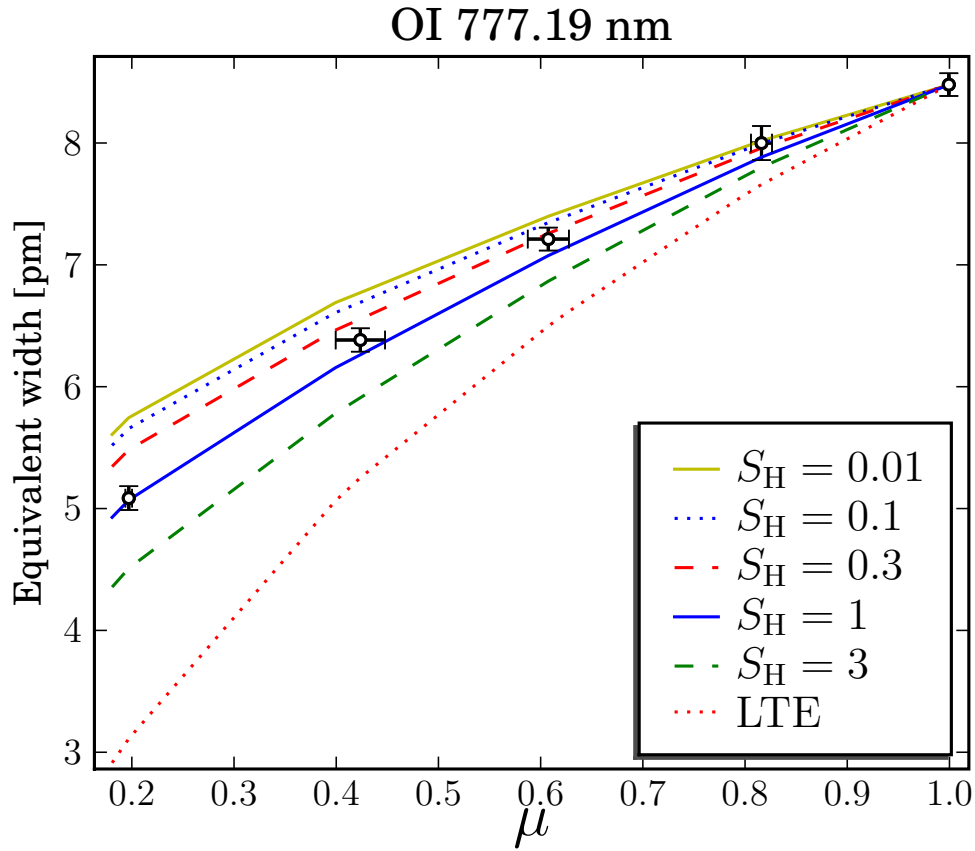,width=15cm}}
\caption{Comparison of the observed (open circles with error bars) and predicted center-to-limb variation of the 
equivalent width of the \oi\ 777.19\,nm triplet line as a function of viewing angle 
($\mu = 1$ corresponds to disk center and 0 to the limb). 
The theoretical curves are the results of full 3D non-LTE line formation calculations
with different adopted efficiency for the poorly known inelastic H collisions; the O abundances
have been normalized to fit the disk-center line strengths in each case. 
The curves are labelled with the scaling factor used to multiply the classical
\citet{1968ZPhy..211..404D} 
formula. The solar observations thus seem to suggest that this recipe
is a rather accurate representation for \oi\
\citep{pereira_clv}. 
\label{f:oiclv}
}
\end{figure}

There are a half-dozen or so permitted, high-excitation \oi\ lines useful for solar abundance purposes.
Of these the relatively strong 777\,nm triplet lines are arguably the most reliable since
the others are either weak or partly blended.
The main challenge when employing the \oi\ lines relates to quantifying the non-LTE effects, which can be significant.
The predicted non-LTE abundance corrections are little dependent on the model atmosphere in question, be it 1D or 3D
\citep{2004A&A...417..751A} 
but they do depend on the adopted collisional cross-sections, in particular the inelastic H collisions
\citep[e.g.][]{1993A&A...275..269K, 2005ARA&A..43..481A}. 
Different approaches have been used in solar abundance works to date where some neglect the H collisions 
altogether based on the available atomic physics data for other elements 
while other use the classical
\citet{1968ZPhy..211..404D} 
formula, possibly with a scaling factor $S_{\rm H}$ that typically varies from 0 to 1.
\citet{2001AIPC..598...23H} 
found $\log \epsilon_{\rm O} = 8.71 \pm 0.05$ using the  
\citet{1974SoPh...39...19H} 
model with granulation corrections estimated from a 2D solar model,
while 
\citet{2004A&A...417..751A} 
obtained a 0.07\,dex lower abundance from a 3D non-LTE study; 
the main difference, however, between the two studies is the adopted H collisions ($S_{\rm H} =1$ and 0, respectively).
\citet{2004A&A...423.1109A} 
attempted to calibrate the poorly known collisional efficiency through a comparison
of the center-to-limb variation of the \oi\ 777\,nm line strengths since the relative importance
of H collisions change as $N_{\rm H}/N_{\rm e}$ increases with smaller optical depths. 
They found that $S_{\rm H} = 1$ was a marginally better fit than $S_{\rm H} = 0$ while
the LTE case could be ruled out at high significance level. 
Recently, 
\citet{pereira_clv} 
have revisited the issue equipped with better solar observations and tested additional $S_{\rm H}$ values
but have arrived at a similar conclusion, as can be seen in Fig. \ref{f:oiclv}.

The \oi -based abundance we adopt here stems from 
\citet{sun_o}, 
which is based on 
full 3D non-LTE calculations with $S_{\rm H} = 1$.
The new electron collisional cross-sections computed by 
\citet{2007A&A...462..781B} 
have also been adopted (see 
\citealt{2009A&A...500.1221F} 
for an account of the impact of these on other stars).
The revised abundance is slightly higher than in 
\citet{2004A&A...417..751A}: 
$\log \epsilon_{\rm O} = 8.69 \pm 0.05$.
Using a different 3D model,
\citet{2008A&A...488.1031C} 
obtained $\log \epsilon_{\rm O} = 8.73 \pm 0.06$ (unweighted mean) for their disk-center \oi\ lines,
in spite of using $S_{\rm H} = 1/3$; with $S_{\rm H} = 1$ their result would be 0.02\,dex higher.
Because they were restricted to 1D non-LTE calculations, they were unable to
carry out line profile fitting as done by 
\citet{sun_o}. 
They were therefore forced to rely on measuring equivalent widths, which
were surprisingly large compared to all previous published studies.
Reassuringly, when the same input data are adopted (e.g. line strengths, H collisional efficiency), the
\citet{2008A&A...488.1031C} 
and
\citet{sun_o} 
analyses are in excellent agreement, which shows that the exact choice of 3D model is of little
consequence for the permitted \oi\ lines.

\citet{1984A&A...141...10G} 
and 
\citet{1984ApJ...282..330S} 
introduced the vibration-rotation and pure rotation lines of OH as solar O abundance indicators.
In their 3D-based analysis,
\citet{2004A&A...417..751A} 
found that the OH lines implied an O abundance only slightly lower than the atomic transitions.
While the vibration lines showed no trends with excitation potential or line strength, the pure
rotation lines did. 
This is not too surprising,
since these strong rotation lines are formed in very high atmospheric layers, which are the most
challenging to model realistically due to for example magnetic fields, non-LTE and non-equilibrium chemistry.
\citet{sun_o} 
revisited the OH lines equipped with a new 3D model, which yielded 
$\log \epsilon_{\rm O} = 8.69$ and 8.69 for the vibration-rotation and pure rotation lines, respectively;
the trend with line strength for the latter were still present but somewhat reduced and now in the opposite direction.
Unfortunately, the OH lines have not yet been analysed using the alternative 3D model of
\citet{2008A&A...488.1031C}. 
\citet{2004ApJ...615.1042M} 
added a few vibration lines from the first overtone band of OH, which we however do not consider
sufficiently reliable abundance indicators due to their extreme weakness. 

The 3D-based solar O abundance that we recommend here,  $\log \epsilon_{\rm O} = 8.69 \pm 0.05$, 
is a mean of the [O\,{\sc i}], O\,{\sc i}, OH vibration-rotation and OH pure rotation results of
\citet{sun_o}. 
The agreement between the different abundance indicators is very satisfactory, which 
is not the case with the
\citet{1974SoPh...39...19H} 
model as seen in Table \ref{t:cno}. 
As explained above, the use of a 3D model is only one factor in the downward
revision of the O abundance from the value $8.83 \pm 0.06$ given in 
\citet{1998SSRv...85..161G} 
with refined non-LTE line formation, better accounting of blends and improved atomic 
and molecular data also playing important roles.

Using CO lines, 
\citet{2006A&A...456..675S} 
determined an isotopic abundance of \linebreak \mbox{$^{16}{\rm O}/^{18}{\rm O} = 479 \pm 29$}.
The significantly lower, non-terrestrial ratio obtained by
\citet{2006ApJS..165..618A} 
is likely the result of their use of 1D model atmospheres. 
Several lines due to $^{12}$C$^{17}$O are present in the solar infrared spectrum 
but due to their extreme weakness we do not believe a trustworthy $^{16}{\rm O}/^{17}{\rm O}$
ratio can be inferred from them at this stage.

\subsection{Intermediate Mass Elements}
\label{s:na-ca}


{\bf Sodium:}
\citet{sun_na-ca} 
based their 3D solar Na abundance determination on five weak \nai\ lines, which all yielded
consistent results ($\sigma = 0.03$\,dex).
The predicted departures from LTE in 1D are slightly negative ($\approx -0.04$\,dex)
and agree well between different studies
\citep{
2003ChJAA...3..316T, 
2004A&A...423..683S, 
2004A&A...423.1109A}. 
The Na abundance we adopt here is slightly larger than the preliminary value presented in
\citet{2005ASPC..336...25A}. 

{\bf Magnesium:}
The abundance of Mg can be inferred from two ionization stages, each with a multitude
of seemingly clean, relatively weak lines
\citep{1978MNRAS.183...79L}. 
Unfortunately, the $gf$-values for \mgi\ and \mgii\ are notoriously uncertain, as
reflected in fairly large line-to-line abundance variations for both species. 
Departures from LTE in 1D line formation by \mgi\ have been considered by
\citet{1998A&A...333..219Z} 
and T. Gehren (2008, private communication);
the abundance corrections are small and positive.
\citet{2004MNRAS.350.1127A} 
and L. Mashonkina (2008, private communication)
have made corresponding calculations for \mgii .
With the predicted non-LTE effects, the mean \mgii\ abundance 
is brought into better agreement with the meteoritic abundance
while \mgi\ is further offset. 
Since we do not have greater confidence in either set of transition probabilities
and non-LTE calculations are available for both species, we simply take a straight mean
of all lines
\citep{sun_na-ca}. 
The attached uncertainty is disappointingly large for such an important element.

{\bf Aluminium:}
Our recommended Al abundance comes from 
\citet{sun_na-ca}, 
who performed an analysis of seven good, weak \ali\ lines with
relatively well-determined $gf$-values.
The departures from LTE are predicted to be small.
We have adopted the non-LTE results from 1D calculations by  
\citet{2004A&A...413.1045G}, 
with the expectation that they will not be very different in 3D.

{\bf Silicon:}
Silicon is a critical element since it provides the absolute normalization for
the meteoritic abundances (Sect. \ref{s:meteorites}).
\citet{2000A&A...359..755A} 
carried out an early 3D-based LTE analysis of \sii\ and \siii\ lines and found
$\log \epsilon_{\rm Si} = 7.51 \pm 0.04$. 
With an improved 3D solar model and non-LTE corrections ($\approx -0.02$\,dex) taken from 
\citet{2008A&A...486..303S}, 
\citet{sun_na-ca} 
found the same value in their reanalysis,  which we adopt here. 
Had we instead opted to use the
\citet{1974SoPh...39...19H} 
1D semi-empirical model the result would have been a mere 0.02\,dex higher.

{\bf Phosphorus:}
\citet{2007A&A...473L...9C} 
carried out a solar P determination using a {\sc co5bold} 3D solar model and found
$\log \epsilon_{\rm P} = 5.46$ $(\sigma=0.04)$ from four \phoi\ lines. 
A slightly lower 3D-based value was obtained by 
\citet{sun_na-ca}, 
who also included one more line:
$\log \epsilon_{\rm P} = 5.41 \pm 0.03$ ($\sigma = 0.03$). 
We adopt this latter value on account of its smaller line-to-line scatter.
No non-LTE investigation has been performed to date for these high-excitation \phoi\ lines.
Departures from LTE are not expected to be significant however, a prediction supported 
by the small non-LTE effects for the similar \si\ lines
\citep{2005PASJ...57..751T}. 

{\bf Sulphur:}
In their 3D-based analysis, 
\citet{2007A&A...470..699C} 
obtained a disturbingly large line-to-line scatter: 
$\log \epsilon_{\rm S} = 7.21 \pm 0.11$ ($\sigma$);
this result includes the non-LTE corrections of  
\citet{2005PASJ...57..751T}. 
Very recently, the issue of the solar S abundance was revisited by 
\citet{sun_na-ca}, 
who included a few additional lines. Their non-LTE abundance is slightly smaller
and, importantly, shows a much reduced uncertainty:  
$\log \epsilon_{\rm S} = 7.12 \pm 0.03$ ($\sigma = 0.03$). 
The differences between 
\citet{sun_na-ca} 
and
\citet{2007A&A...470..699C} 
can be partly traced to the adopted equivalent widths, but this cannot be the whole explanation. 
Interestingly, 
\citet{sun_na-ca} 
concluded that the forbidden [\si ] 1082\,nm line returns too high an abundance by $0.3$\,dex, 
which may signal an unidentified blend, non-LTE effects or that the theoretical transition probability is in error.
\citet{2007A&A...467L..11C} 
on the other hand found excellent agreement with the meteoritic value for this line.
They adopted however an outdated $gf$-value and a much smaller equivalent width, which
we are unable to reconcile with the observed solar spectrum. 

{\bf Potassium:}
\citet{sun_na-ca} 
based their 3D analysis on
six \ki\ lines, including the strong resonance line at 769.9\,nm. 
As in previous studies, the line-to-line scatter is disappointingly large for these lines, even
when taking into account the theoretical non-LTE effects computed by 
\citet{2006A&A...453..723Z}. 
In terms of abundances, the non-LTE corrections are substantial for the two strongest
lines ($\ge -0.2$\,dex). 

{\bf Calcium:}
The solar Ca abundance can be determined from both
\cai\ and \caii\ 
\citep{1978MNRAS.183...79L}, 
although there are problems with both species:
most of the \cai\ lines are rather strong, whilst the uncertainties in the
transition probabilities for the \caii\ lines are still unsatisfactory large
(Opacity Project data assuming LS-coupling).
\citet{sun_na-ca} 
found good agreement between the two ionization stages and obtained
a final Ca abundance of 
$\log \epsilon_{\rm Ca} = 6.34 \pm 0.04$ ($\sigma = 0.04$) from all \cai\ and \caii\ lines, including
1D non-LTE abundance corrections computed by
\citet{2007A&A...461..261M}. 
We note that the forbidden [\caii ] line at 732.3\,nm 
implies an abundance which is fully consistent with the other transitions.

\subsection{Iron-peak Elements}
\label{s:fepeak}


{\bf Scandium:} 
With its relatively low ionization potential, Sc is prone to over-ionization via non-LTE effects
\citep{2008A&A...481..489Z}. 
The impact on abundances from \sci\ lines is dramatic ($+0.15$\,dex relative to LTE on 
average for the lines considered by \citealt{sun_fepeak}), but predictably  
mild for \scii\ ($-0.01$\,dex).  Good lines, oscillator strengths and hyperfine constants
are available for both ionization stages.
With five weak lines of \sci, twelve of \scii\
and the non-LTE corrections of
\citet{2008A&A...481..489Z}, 
\citet{sun_fepeak}
have produced 3D-based abundances showing very good agreement between the two ionization stages.
In Table \ref{t:sun} we give the mean of the abundances from all lines of both species, 
where \scii\ ultimately dominates because of the greater number of lines.  
The $+0.10$\,dex difference between the new Sc abundance and the meteoritic one is intriguing.

{\bf Titanium:} 
Very many good lines of \tii\ are present in the Sun, virtually all with excellent atomic data.  
Whilst fewer lines of \tiii\ are available and the atomic data are not quite as good as for \tii, 
the situation is still relatively acceptable.  The one glaring absence is a study of non-LTE effects for this element.  
As is the case for its neighbours Sc and V, the low ionization potential of Ti and the 
resulting minority status of \tii\ are expected to significantly skew the abundances it returns 
under the assumption of LTE.  Using the improved 3D model atmosphere,  
\citet{sun_fepeak} 
found stark disagreement between LTE abundances from \tii\ and \tiii\  
(\tii: $\log \epsilon_{\rm Ti} = 4.85\pm0.06$, \tiii: $\log \epsilon_{\rm Ti} = 4.99\pm0.04$), 
though with very low scatter from each species individually ($\sigma = 0.03$ for both stages).  
In the absence of any non-LTE study, we favour \tiii\ in Table~\ref{t:sun} by taking the mean of the two 
results weighted according to their total uncertainties; 
we only retain any weighting for the \tii\ result because of the sheer number of 
excellent lines and oscillator strengths it includes. 

{\bf Vanadium:}
Neutral vanadium has scores of excellent lines in the visible solar spectrum, 
whilst very few are available from \vii, all extremely perturbed.  
On the other hand, \vi\ is expected to show similarly strong non-LTE effects 
as verified for \sci\ and suspected for \tii, owing to the elements' similar potential for over-ionization.  
The impact of non-LTE line formation on V has not yet been the subject of a dedicated study. 
\citet{sun_fepeak} 
obtain a 3D LTE abundance of $\log \epsilon_{\rm V} = 3.83$ 
from \vi\ lines.
\citeauthor{sun_fepeak}'s analysis of \vii\ lines yields an abundance quite consistent with this corrected value, 
but the poor quality of the lines employed makes the result inconclusive. 
In the absence of any good results for \vii , 
we adopt the \vi\  value but adjust it by the mean non-LTE effect seen in \sci\ and \cri\ ($+0.10$\,dex).  

{\bf Chromium:}
The solar spectrum exhibits a large number of clean, weak lines of both neutral and once-ionised chromium.  
Unfortunately, no reliable oscillator strengths are available for \crii.  
Neither is any estimate of the impact of non-LTE effects on Cr available; 
given its relatively low ionization potential, Cr should consist mostly of \crii\ in the Sun, 
possibly leading to non-LTE effects for \cri. 
Recent non-LTE calculations by M. Bergemann (private communication) confirm departures from LTE at
the level of $+0.04$\,dex.
Using their own measured oscillator strengths for \cri, 
\citet{2007ApJ...667.1267S} 
recently revised the solar chromium abundance.  This result was updated by
\citet{sun_fepeak} 
using a 3D model.
Given the much greater number of \cri\ lines in the sample, and the poor quality of the \crii\ oscillator strengths, 
we allow \cri\ to dominate the value in Table~\ref{t:sun} 
by taking the mean of abundances from all \cri\ and \crii\ lines together.  
The only reason we allow abundances from \crii\ lines to enter into the 
mean is our suspicion of an as yet unquantified non-LTE effect for \cri.

{\bf Manganese:}
Following years of confusion now traceable mostly to poor oscillator strengths, 
the solar manganese abundance was recently rederived by 
\citet{2007A&A...472L..43B}, 
using their own improved \mni\ $gf$-values, 
extensive hyperfine splitting data and the non-LTE calculations of
\citet{2007A&A...473..291B}. 
This result was updated by
\citet{sun_fepeak} 
using a 3D solar model, bringing the photospheric Mn abundance even 
closer to the meteoritic value.  
\mni\ shows moderate non-LTE effects (about $+0.06$\,dex), 
consistent with its ionization potential being intermediate between the lower values 
seen in the lighter members of the iron group Sc, Ti, V and Cr 
and the higher values exhibited by the heavier elements Fe, Co and Ni. 

{\bf Iron:}
In many circumstances iron serves as a proxy for the overall
metal content and as a reference against which other elemental abundances 
are measured.
An intense debate long raged as to whether the solar Fe abundance inferred from
\fei\ lines should be
$\log \epsilon_{\rm Fe} \approx 7.7$
\citep[e.g.][]{1995A&A...296..217B} 
or $\log \epsilon_{\rm Fe} \approx 7.5$
\citep[e.g.][]{1995A&A...296..233H}, 
with the choice of equivalent widths, $gf$-values, microturbulence parameters and
pressure-damping constants all playing a role. 
The advent of improved transition probabilities for \feii\ lines
\citep[e.g.][]{
1992A&A...259..301H, 
1999A&A...342..610S} 
and a proper treatment of pressure broadening
\citep[e.g.][]{
1997MNRAS.284..202A, 
2000A&AS..142..467B} 
strengthened the case for a low Fe abundance. An unsatisfactory combination of large abundance scatter, 
the presence of a free parameter (microturbulence), and a trend with excitation potential still remained 
with the 
\citet{1974SoPh...39...19H} 
model, however
\citep{1999A&A...347..348G}. 
Using a 3D hydrodynamical solar model and LTE line profile fitting, 
\citet{2000A&A...359..743A} 
succeeded in achieving consistent results for \fei\ and \feii\ lines without invoking any microturbulence:
$\log \epsilon_{\rm Fe} = 7.45\pm0.05$.

\begin{figure}
\centerline{\psfig{figure=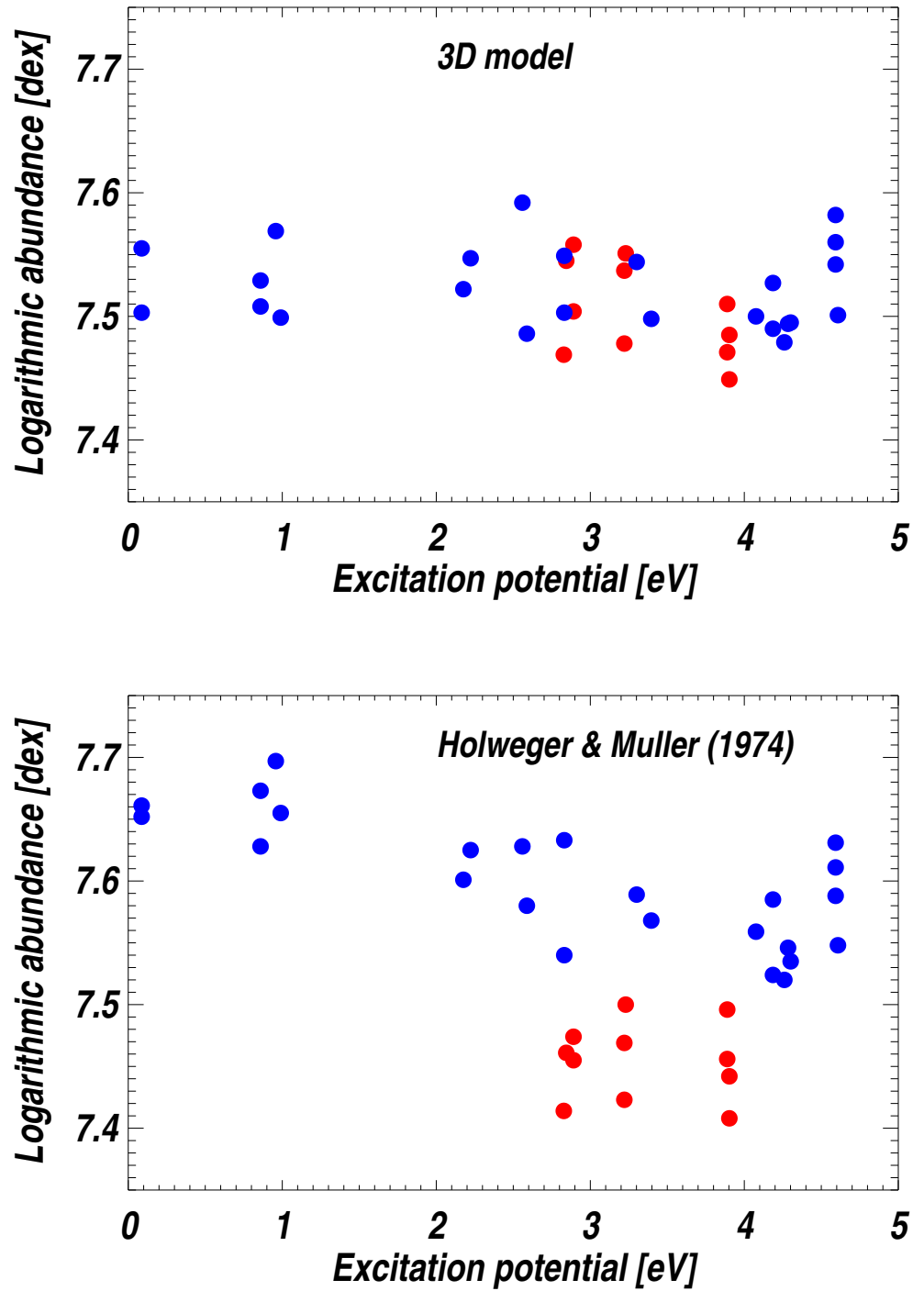,width=15cm}}
\caption{{\em Upper panel:} The derived \fei\ (blue circles) and \feii\ (red circles) abundances 
\citep{sun_fepeak} 
as a function of excitation potential of the lower level
when using a 3D hydrodynamical solar model atmosphere. 
{\em Lower panel:} Same as above but with the 1D semi-empirical 
\citet{1974SoPh...39...19H} 
model. Note the offset between \fei\ and \feii\ and the trend with excitation
potential for the \fei\ lines, neither of which are present when employing the 3D model.
\label{f:fe}
}
\end{figure}

The solar Fe abundance we advocate here comes from 
\citet{sun_fepeak}.  
The \fei -based abundance of 
$\log \epsilon_{\rm Fe} \approx 7.52\pm0.05$
from weak lines is slightly higher than in 
\citet{2000A&A...359..743A} 
on account of a slightly shallower mean temperature structure in the new 3D model and allowance
of departures from LTE. 
The non-LTE abundance corrections have been estimated to be $+0.03$\,dex for \fei\ 
\citep{2003A&A...407..691K, 
2005A&A...442..643C}, 
although this depends on the still poorly-known 
inelastic H collisions 
\citep[see][ and references therein]{2005ARA&A..43..481A}. 
In fact, a more reliable abundance now comes from \feii\ lines, thanks to the 
improved transition probabilities of
\citet{2009A&A...497..611M} 
and the broadening treatment of
\citet{2005A&A...435..373B}. 
Together these imply
$\log \epsilon_{\rm Fe} \approx 7.50\pm0.04 (\sigma = 0.04)$
with the same 3D model. 
We note that with the 
\citet{1974SoPh...39...19H} 
model, there is a trend in \fei\ abundances with excitation potential and
a discrepancy of 0.14\,dex between \fei\ and \feii.  Both have disappeared 
with the 3D solar model (Fig. \ref{f:fe}).
Reassuringly, the situation regarding Fe has greatly improved over the past decade. 

{\bf Cobalt:}
The fairly high ionization potential of Co and the availability of many good 
\coi\ lines, $gf$-values and hyperfine constants should make its solar analysis quite straightforward.  
\citet{sun_fepeak} 
found a 3D LTE abundance of 
$\log \epsilon_{\rm Co} = 4.88$, 
in agreement with the meteoritic value.  
Taking into account the surprisingly large non-LTE corrections of
\citet{2008PhST..133a4013B} 
changes the final abundance to $\log \epsilon_{\rm Co} = 4.99\pm0.07$ ($\sigma = 0.05$).

{\bf Nickel:}
The solar Ni abundance was estimated by  
\citet{2009ApJ...691L.119S} 
from 17 weak \nii\ lines, showing a very small abundance scatter. 
The value given in Table \ref{t:sun} is a slight revision of this result
\citep{sun_fepeak} 
based on the improved 3D model atmosphere that has been employed throughout this review.

\subsection{Neutron Capture Elements}
\label{s:neutroncapture}


Quite some work on the solar abundances of the heavy elements has been 
carried out over the past decade, largely driven by improvements in
atomic transition probabilities. 
In this context, we would like to particularly highlight the very careful laboratory
measurements performed by the Wisconsin group
\citep[e.g.][]{2009ApJS..182...51L}. 
In a beautiful series of papers, they have systematically and methodically measured 
radiative lifetimes, branching fractions, oscillator strengths, isotopic shifts and hyperfine constants
for all the rare Earth elements:
La \citep{2001ApJ...556..452L}, 
Ce \citep{2009ApJS..182...51L}, 
Pr  \citep{2009ApJS..182...80S}, 
Nd \citep{2003ApJS..148..543D}, 
Sm \citep{2006ApJS..162..227L}, 
Eu \citep{2001ApJ...563.1075L}, 
Gd \citep{2006ApJS..167..292D}, 
Tb  \citep{2001ApJS..137..341L}, 
Dy  \citep{2009ApJS..182...80S}, 
Ho \citep{2004ApJ...604..850L}, 
Er \citep{2008ApJS..178...71L}, 
Tm, Yb and Lu  \citep{2009ApJS..182...80S}; 
the group has also carried out a similar study for
Hf \citep{2007ApJS..169..120L}. 
In their solar abundance analyses,
they have employed the 1D
\citet{1974SoPh...39...19H} 
model atmosphere and assumed LTE for the spectrum synthesis.
Although all the utilized transitions are from
the majority ionization stage (once ionized) for the rare Earths, one still 
expects a minor temperature and thus model-atmosphere dependence,
since the lines originate from atomic levels with low excitation potentials ($\chi_{\rm exc} \le 1.5$\,eV). 
The recommended solar abundances provided in Table \ref{t:sun}
for these elements have been modified from the original sources by 
calculating the abundance difference between a 3D hydrodynamical model atmosphere and the 
\citet{1974SoPh...39...19H} 
model, each computed using the same line formation code with consistent input data
\citep{sun_heavy}. 
For the majority of these lines, such (3D-HM) abundance
corrections amount to $\approx -0.04$\,dex.
In addition, the 1D non-LTE effects on \euii\ predicted by
\citet{2000A&A...364..249M} 
have been taken into account in the 3D result; non-LTE studies are not available for
any other rare Earth elements.

A similar procedure has been implemented for 
Cd \citep{1990A&A...239..367Y}, 
W \citep{1982SoPh...81....3H}, 
Os \citep{2006A&A...448.1207Q}, 
Ir \citep{1988A&A...203..378Y} 
and
Au \citep{1986A&A...164..395Y}, 
all of which have only heavily blended transitions available in the solar spectrum, necessitating 
careful and time-consuming spectrum syntheses. 
Due to blends, we consider the two As\,{\sc i} lines employed by 
\citet{2001KFNT...17...37G} 
too unreliable to base a meaningful abundance determination on
and consequently do not provide a solar photospheric As abundance here. 
Similarly, we are unable to defend an estimate of the Sb abundance from the 
very weak and heavily perturbed, purported Sb\,{\sc i} line at 323.2\,nm reported by 
\citet{1976Sci...191.1223R}. 

For all other neutron capture elements, the adopted abundances come from
a complete 3D-based reanalysis 
using the best atomic data
and newly measured equivalent widths or spectrum synthesis for the most reliable lines.
\citet{sun_heavy} 
present the solar analyses of 
Cu, Zn, Ga, Ge, Rb, Sr, Y, Zr, Nb, Mo, Ru, Rh, Pd, Ag, Cd, Sn, Ba, Pb
and Th, which we adopt throughout;
the case of Ru is also discussed in 
\citet{2009MNRAS.396.2124F}. 
The available 1D non-LTE abundance corrections for \zni\
\citep{2005PASJ...57..751T}, 
\srii\ 
\citep{2001A&A...376..232M} 
and \baii\
\citep{2000A&A...364..249M} 
have been included in these estimates.
For comparison purposes we have also repeated the line calculations with
an assortment of 1D model atmospheres. We have found very good agreement ($<0.02$\,dex)
with recent studies using the 
\citet{1974SoPh...39...19H} 
model when allowance is made for differences in adopted transition probabilities
and equivalent widths for 
Ge \citep{1999MNRAS.303..721B}, 
Sr \citep{2000MNRAS.311..535B}, 
Pd \citep{2006A&A...452..357X} 
and Pb \citep{2000MNRAS.312..116B}. 

A handful of elements have previously been exposed to a 3D-based solar
abundance analysis.
\citet{2006A&A...456.1181L} 
made use of new transition probabilities for Zr\,{\sc ii} and employed the 3D solar model of 
\citet{2000A&A...359..729A}. 
Not surprisingly, the agreement with 
\citet{sun_heavy} 
is very good.
Eu has been the focus of
\citet{2008A&A...484..841M}, 
who employed a 3D {\sc co5bold} hydrodynamical model and 
found a mean Eu abundance in perfect agreement with ours.
The 3D LTE abundance for Hf from 
\citet{2008A&A...483..591C} 
is only 0.02\,dex higher than the corresponding value by 
\citet{sun_heavy}, 
who corrected the spectrum synthesis results of 
\citet{2007ApJS..169..120L} 
for 3D effects rather than attempting to measure the equivalent widths of these weak and blended Hf\,{\sc ii} lines. 
\citet{2008A&A...483..591C} 
also performed a detailed investigation using 3D line profile fitting of the 
Th\,{\sc ii} 401.9\,nm line, which is distorted by two \coi\ and \vi\ lines (not \vii\ as stated in
their paper), as well as being in the red wing of a relatively strong feature from \fei\ and \nii .
Interestingly, they argued that it is important to take into account the line asymmetry of
the  \fei+\nii\ blend introduced by the convective motions
\citep[e.g.][]{2000A&A...359..729A}, 
which otherwise would lead to an overestimation of the Th abundance by $\approx 0.1$\,dex. 
Unfortunately 
\citet{2008A&A...483..591C} 
did not specify what abundances they used for the blending lines, so
it is difficult to exactly reproduce their result; the value obtained by 
\citet{sun_heavy} 
is consistent with that of \citeauthor{2008A&A...483..591C}, though smaller by $0.06$\,dex.

\subsection{Abundances from Sunspots: F, Cl, In, Tl}
\label{s:sunspot}

The photospheric abundances of a few elements are not possible to measure using
spectroscopy of the quiet Sun. The solar fluorine and chlorine abundances 
have instead been inferred from the infrared spectrum of sunspots using
lines of HF  
\citep{1969ApL.....4..143H} 
and HCl
\citep{1972ApJ...175L..95H}. 
Given the improvements in molecular data, observational quality and sunspot modelling 
in the intervening years since the original workds, 
it seems like a critical re-evaluation of these studies is warranted but not attempted here. 
The same holds for Tl 
\citep{1972SoPh...26..250L}. 
We note that in the meantime a more reliable estimate of the solar Cl may be provided by the abundance
of nearby H\,{\sc ii} regions, which is 
$\log \epsilon_{\rm Cl} = 5.32 \pm 0.07$
\citep{ 2007ApJ...670..457G}. 
Recently, 
\citet{2008MNRAS.384..370V} 
have revisited the solar In abundance. 
In contrast to previous studies based on the quiet Sun, they find an abundance from a sunspot spectrum 
in agreement with the meteoritic value.

\subsection{Indirect Determinations: Noble Gases}
\label{s:noble}


Due to their high excitation potentials, there are no photospheric lines
of the noble gases helium, neon, argon, krypton and xenon available in
the solar spectrum, which forces one to estimate the solar abundances of these elements 
through indirect methods.  

{\bf Helium:} 
A highly accurate He abundance determination is possible through helioseismology. 
The technique utilizes the change in the adiabatic index $\Gamma_1$ in the \heii\ ionization zone that
takes place at a solar radius of $r/R_\odot \approx 0.98$. The result is quite insensitive to
the reference solar model employed for the inversion, but does depend on the adopted
equation-of-state. 
\citet{2004ApJ...606L..85B} 
found a He mass fraction in the convection zone of $Y_{\rm s} = 0.2485 \pm 0.0034$, which
corresponds to a He abundance by number of $\log \epsilon_{\rm He} = 10.93 \pm 0.01$.
The uncertainty essentially reflects the difference between the MHD 
\citep{1988ApJ...331..815M} 
and OPAL
\citep{2002ApJ...576.1064R} 
equation-of-state, but there is also a minor dependence on the assumed metallicity of the
reference solar model for the inversion
\citep{2006ApJ...646..560T}. 
A similar value is obtained when calibrating solar interior models to achieve the correct mass, luminosity
and radius at the present solar age, if the chemical composition of 
\citet{1998SSRv...85..161G} 
is adopted and diffusion is considered. 
With the solar abundances of 
\citet{2005ASPC..336...25A} 
on the other hand, the calibrated proto-solar He abundance would imply
a surface value lower than the helioseismic measurement by 0.02\,dex.
Such models are also in conflict with other helioseismological evidence such as 
the depth of the convection zone and the sound speed variation with depth,
as discussed in Sect. \ref{s:helioseismology}.

{\bf Neon:} 
The Ne abundance can be inferred from X-ray and UV spectroscopy of
the solar corona and solar flares as well as directly from the solar wind. 
The interpretation is, however, complicated by the so-called first ionization potential (FIP) effect: 
elements with ionization potential $\chi_{\rm ion} < 10$\,eV are enhanced in 
the upper solar atmosphere and solar wind compared with their photospheric values,
whilst the high ionization elements are not or only slightly affected
\citep{2003SSRv..107..665F}. 
Rather than measuring the Ne abundance directly, a common approach is to determine
the abundance relative to a reference element and assume that
the ratio is the same in the photosphere. 
Here O is the reference element of choice, although we note that there
is still a significant difference in ionization potential between O (13.6\,eV) and Ne (21.6\,eV), which could
potentially introduce a systematic error given our poor understanding of the physical reasons for the FIP effect.
The degree of chemical separation varies significantly, being more severe in regions of higher solar activity. 
\citet{2005A&A...444L..45Y} 
found a Ne/O ratio of $0.175\pm0.031$ for the quiet Sun, which
is the value we adopt here to convert our photospheric O abundance to a Ne content.
This result is particularly appealing for our purposes since
\citet{2005A&A...439..361Y} 
has demonstrated that the average quiet Sun does not show any FIP effect, a conclusion
which is further corroborated when using the solar abundances given herein.
This ratio is consistent with the results of 
\citet{2005ApJ...634L.197S} 
for active regions ($0.175\pm0.074$) and for the solar wind ($0.14\pm0.03$) by
\citet{2007A&A...471..315B}. 
The corresponding ratio in solar energetic particles is very similar and remarkably constant: $0.152\pm0.006$ 
\citep{1999SSRv...90..413R}, 
which was  adopted by 
\citet{2005ASPC..336...25A}. 
In spite of its very small uncertainty, here we weight this result less
than from the quiet Sun due to the potential modification through the FIP effect. 
Together with our recommended O abundance,  the photospheric Ne abundance becomes
$\log \epsilon_{\rm Ne} = 7.93 \pm 0.09$.

Our value is smaller than the solar flare determination ($\log \epsilon_{\rm Ne} = 8.11 \pm 0.12$) by 
\citet{2007ApJ...659..743L}, 
but we consider the quiet Sun a more reliable indicator than solar flares due to possible fractionation processes
in the latter.
While the Ne abundance  ($\log \epsilon_{\rm Ne} = 7.96 \pm 0.13$)  of 
\citet{2007A&A...471..315B} 
agrees with ours, there are significant additional uncertainties in this result on account of using
He as a reference element, as it is known to be highly variable in the solar wind. 

\citet{2005Natur.436..525D} 
have argued on the basis of X-ray spectroscopy of nearby stars that the 
Ne/O ratio is significantly higher (0.41) than the solar ratios discussed above.
Their sample is, however, heavily biased towards active stars, which
experience an inverse FIP effect:
elements with $\chi_{\rm ion} > 10$\,eV are enhanced in the corona compared with
low ionization potential species
\citep{2004A&ARv..12...71G}. 
Indeed, 
\citet{2008A&A...486..995R} 
show that there is a systematic trend in Ne/O with activity level such
that other inactive stars have similar low ratios to the Sun.
One would therefore suspect that the high coronal Ne/O ratio inferred by  
\citet{2005Natur.436..525D} 
is a consequence of the difference in ionization potential, 
making Ne more susceptible to fractionation than O in high activity environments.
As will be further discussed in Sect. \ref{s:solarneighborhood}, a more reliable stellar gauge of
the solar Ne abundance is provided by OB stars in the solar neighborhood. 
Both the Ne/O ratio ($0.20\pm0.03$) and the absolute Ne abundance 
($\log \epsilon_{\rm Ne} = 8.08 \pm 0.03$) estimated in this way are in very good 
agreement with the solar values we adopt here when the effects of solar
diffusion and Galactic chemical enrichment over the past 4.5\,Gyr are accounted for
\citep{2008ApJ...688L.103P}. 

{\bf Argon:}  
As discussed in detail by
\citet{2008ApJ...674..607L}, 
there are a wide range of methods for inferring the solar Ar abundance: solar wind measurements,
solar flares and energetic particles, nuclear statistical
equilibrium with the abundances of the nearby $\alpha$-elements $^{28}$S and $^{40}$Ca,
and comparison with other solar system objects (Jupiter) and the solar neighborhood 
(B stars, planetary nebulae, H\,{\sc ii} regions).
We follow the procedure of 
\citet{2008ApJ...674..607L} 
but employ our preferred values for the photospheric abundances of O (for the solar wind, flares and energetic
particles), Kr and Xe (for analyses of Jupiter), and the lower He abundance appropriate for the upper atmosphere
instead of the photospheric value (for solar wind analyses). 
The quoted errors for all methods in the literature 
are only the dispersions of the measurements, so 
likely underestimate of the true uncertainties. 
We therefore take a straight unweighted mean of all techniques, which leads to
$\log \epsilon_{\rm Ar} = 6.40 \pm 0.13$.
Our value is 0.1\,dex lower than the result of
\citet{2008ApJ...674..607L} 
but 0.2\,dex higher than 
\citet{2005ASPC..336...25A}, 
which was based only on the Ar/O ratio in solar energetic particles
\citep{1999SSRv...90..413R}. 

{\bf Krypton:} 
The Kr abundance has been estimated from interpolation of the theoretical
$s$-process production rates, since the cross-sections for
neutron capture of nearby pure $s$-isotopes have been accurately measured experimentally
(Palme \& Beer 1993). 
Our value accounts for a slight revision of the photospheric Si abundance, which serves
as a normalization of the $s$-process predictions. 
This Kr estimate agrees well with the value inferred from the solar wind Kr/O ratio together
with our photospheric O abundance
\citep{1994RSPTA.349..213G, 2008AGUFM.P42A..08H}. 

{\bf Xenon:} 
The procedure to estimate the Xe content is the same as for Kr
\citep{2005ASPC..336..255K}. 
For Xe, however, this $s$-process estimated value is 0.4\,dex lower than deduced from the solar wind as
measured from lunar regoliths
\citep{1994RSPTA.349..213G} 
and the Genesis space probe
\citep{2008AGUFM.P42A..08H}. 
This may suggest that Xe has experienced fractionation 
through the FIP effect in spite of its relatively high ionization energy 
($\chi_{\rm ion} = 12.1$\,eV).

\subsection{Solar Isotopic Abundances}
\label{s:isotopes}

As a convenience to the reader, in Table \ref{t:isotopes} we give our best estimates
of the proto-solar isotopic fractions for each element. 
In most cases the terrestrial isotopic abundance ratios as recommended by International Union
of Pure and Applied Chemistry 
\citep{1998JPCRD..27.1275R} 
have been adopted as representative of the solar system values;
direct solar isotopic information can only be gleaned from CO vibration-rotation lines
in the infrared spectrum of the Sun
\citep[e.g.][]{2006A&A...456..675S}, 
although with comparatively large error bars by terrestrial standards.
It should be noted that except for C and O,
the isotopic fractions given here are the more conservative ``representative isotopic composition''
from the compilation of 
\citet{1998JPCRD..27.1275R}, 
in contrast to 
\citet{2003ApJ...591.1220L} 
who opted for the ``best measurement from a single terrestrial source'' from the same reference. 
For most volatile elements more reliable isotopic
fractions are available from alternative solar system sources due to strong
depletion of terrestrial samples, as discussed below.

The proto-solar deuterium abundance can be estimated from observations of Jupiter.
Using ISO spectra of the Jovian atmosphere and taking into account an expected
$5-10$\% enrichment during the planet's formation,
\citet{2001A&A...370..610L} 
obtained a proto-solar abundance of ${\rm D/H} = (2.1 \pm 0.4) \cdot 10^{-5}$.
An alternative approach makes use of the measured enrichment of the
$^3{\rm He}/^4{\rm He}$ ratio in the solar wind in comparison with the proto-solar ratio 
$^3{\rm He}/^4{\rm He} = (1.66 \pm 0.05) \cdot 10^{-4}$ inferred from Jupiter
\citep{1998SSRv...84..251M}, 
since the additional $^3$He can be attributed to D-burning in the Sun
\citep{2000IAUS..198..224G}. 
Together with the solar wind ratio $^3{\rm He}/^4{\rm He} = (4.53 \pm 0.03) \cdot 10^{-4}$
from Genesis measurements 
\citep{2008LPI....39.1779H}, 
the Jovian ratio implies ${\rm D/H} = (1.96 \pm 0.3) \cdot 10^{-5}$;
most of the uncertainty comes from the poorly-known He fractionation in the solar wind.
From the two methods, we finally estimate a proto-solar abundance of 
${\rm D/H} = (2.0 \pm 0.2) \cdot 10^{-5}$.
This is consistent with the present-day local interstellar medium abundance
\citep{2006ApJ...647.1106L}, 
which implies only a small degree of astration over the past 4.5\,Gyr.

The $^{12}{\rm C}/^{13}{\rm C}$ ratio does not appear to vary significantly between different
solar system sources, so we simply adopt the accurate terrestrial reference value of Vienna Peeddee belemnite:
$^{12}{\rm C}/^{13}{\rm C} = 89.4 \pm 0.2$
\citep{Coplen_IUPAC02}. 
We note that the ``representative isotopic composition'' from 
\citet{1998JPCRD..27.1275R} 
corresponds to the significantly more uncertain value $^{12}{\rm C}/^{13}{\rm C} = 92 \pm 7$.
The ratio inferred from the solar photosphere using CO lines ($86.8 \pm 3.8$)
is consistent with the terrestrial value
\citep{2006A&A...456..675S}. 

The  $^{14}{\rm N}/^{15}{\rm N}$ ratio varies tremendously between various solar
system objects, from 100 to 450
\citep{2001ApJ...553L..77O}. 
Here we adopt the Jovian value $435 \pm 57$ as representative of the proto-solar ratio,
in agreement with values in the interstellar medium and 
carbonaceous chondrites 
\citep{2007ApJ...656L..33M}. 

Oxygen isotopic abundances display a variability of up to a few percent
within the solar system. 
The best values to adopt for the proto-solar oxygen isotopic ratios are still debated
\citep[e.g.][]{2006Natur.440..776I, 2005Natur.434..619H}. 
The first results from the Genesis mission should appear very soon
\citep{2008AGUFM.P42A..07M}, 
which will hopefully shed some light on the issue.
Until such further clarification, we adopt the terrestrial values represented in 
Vienna Standard Mean Ocean Water: 
$^{16}{\rm O}/^{18}{\rm O} = 498.7 \pm 0.1$ and
$^{16}{\rm O}/^{17}{\rm O} = 2632 \pm 7$
\citep{Coplen_IUPAC02}. 
The solar photospheric ratio $^{16}{\rm O}/^{18}{\rm O} = 479 \pm 29$ obtained by 
\citet{2006A&A...456..675S} 
is consistent with the terrestrial value, but with larger uncertainties
(although in fairness the terrestrial ``representative isotopic composition'' from 
\citet{1998JPCRD..27.1275R} 
is equally uncertain: $487 \pm 35$).
Surprisingly, 
\citet{2006ApJS..165..618A} 
found strongly sub-terrestrial ratios
(e.g.~$^{16}{\rm O}/^{18}{\rm O} = 440 \pm 20$) from their analysis of the solar CO lines, 
probably due to an inadequate consideration of the differing
temperature sensitivities of the isotopomer transitions with 1D model atmospheres
\citep{2006A&A...456..675S}. 
Lines from $^{12}{\rm C}^{17}{\rm O}$ are also visible in the solar infrared spectrum, but
due to their extreme weakness their reliability as abundance indicators is questionable. 

For Ne, Ar, Kr and Xe we adopt the recommended values of 
\citet{wieler02} 
derived from solar wind and lunar sample analyses. 
Very similar ratios have recently been presented from preliminary studies
of solar wind samples from the Genesis mission
\citep{2008AGUFM.P42A..08H}. 
For Mo, Dy, Er, Yb and Lu we have exchanged the 
\citet{1998JPCRD..27.1275R} 
values with the more recent measurements given in 
\citet{2009arXiv0901.1149L}. 
Similarly the proto-solar values given here 
for the radioactive nuclides have been taken from the same source.


\subsection{Conversion From Photospheric to Bulk Abundances}
\label{s:bulk}


The elemental abundances given in Table \ref{t:sun} are those currently present in
the solar photosphere and by extension the whole convection zone. 
Due to the combined effects of
thermal diffusion, gravitational settling and radiative acceleration (often referred to collectively as
diffusion) over the past
4.56\,Gyr, the values in Table~\ref{t:sun} differ slightly from the proto-solar chemical composition. 
In comparison with present-day photospheric values, the bulk composition of the Sun is higher by 0.05\,dex for helium and 
0.04\,dex for the heavier elements
\citep{2002JGRA..107.1442T}. 
The uncertainty in the overall diffusion correction is judged to be about 0.01\,dex. 
The element-to-element variation in the expected diffusion is very small ($<0.01$\,dex) and therefore
currently not observationally testable by a comparison of photospheric
and meteoritic abundances.
%
We note that 
\citet{2003ApJ...591.1220L} 
adopted a somewhat larger correction, 0.074\,dex, than that of 
\citet{2002JGRA..107.1442T}. 
She based her estimate on the results of 
\citet{2003ApJ...583.1004B}, 
who computed their solar models using a less detailed diffusion model.

\subsection{Solar metallicity}
\label{s:metallicity}


With the solar chemical composition we recommend here, the mass fractions of
H, He and metals in the present-day photosphere become $X=0.7381$, $Y=0.2485 $ and $Z=0.0134$.
The respective values for the bulk composition are $0.7154$, $0.2703$ and $0.0142$. 
In other words, the solar metallicity is no longer the canonical $2$\% recommended by
\citet{1989GeCoA..53..197A} 
but rather a substantially smaller $1.4$\%. 
Indeed, the intervening years have seen a steady decrease in $Z/X$ from $0.027$
to our preferred value of $0.018$.
This is illustrated in Table \ref{t:metallicity}, which collects the resulting
mass fractions from a number of widely used compilations of the solar chemical composition
over the past two decades.
Our metallicity reverses this declining trend of $Z$ with time by being 
slightly larger than advocated by \citet{2005ASPC..336...25A},
which is mainly the result of the somewhat higher C, O, Ne and Fe abundances.

For completeness, we note that the He abundances recommended by 
\citet{1989GeCoA..53..197A} 
and
\citet{1993oee..conf...15G} 
were based on calibration of solar models and hence corresponded to bulk composition; 
here we have substituted those with 
the helioseismic convection zone He content (Sect. \ref{s:noble}) when computing the present-day photospheric values.
Finally, from Table \ref{t:metallicity} it is clear that the surface He abundance employed by 
\citet{2003ApJ...591.1220L} 
is too low on account of adopting a too efficient elemental diffusion, as discussed in Sect. \ref{s:bulk}.

\section{OTHER METHODS FOR INFERRING SOLAR ABUNDANCES}
\label{s:comparison}

\subsection{Meteorites}
\label{s:meteorites}


Meteoritic data on the elemental abundances constitute an excellent comparator for 
the results from the solar photosphere presented above. Not all meteorites are suitable
for the purpose though; only a very small subsample are believed to
accurately reflect the chemical composition of the proto-solar nebula, since the majority
have undergone varying degrees of fractionation and alteration. 
The most primitive, undifferentiated meteorites in this respect 
are the so-called CI (or C1) carbonaceous chondrites, of which there are only five known:
the Alais, Ivuna, Orgueil, Revelstoke and Tonk meteorites
\citep{2003ApJ...591.1220L}. 
Of these, Orgueil is the by far largest and has therefore been exposed to the greatest number of analyses.
It should be noted that while CI chondrites are believed not to have been noticeably chemically
fractionated, they have still been severely modified mineralogically by aqueous alteration and
thermal metamorphism. 
The main advantage of meteorites as a cosmic abundance standard is the
extraordinarily high precision obtained, even for isotopic abundances, through mass spectroscopy.
It comes at a cost, however, since the volatile elements H, C, N, O and the
noble gases -- which also happen to be the most abundant elements -- have been severely depleted.
As a consequence, the abundances in meteorites are pegged to the
Si content, rather than H as is the case in solar and stellar spectroscopy.  

\begin{figure}
\centerline{\psfig{figure=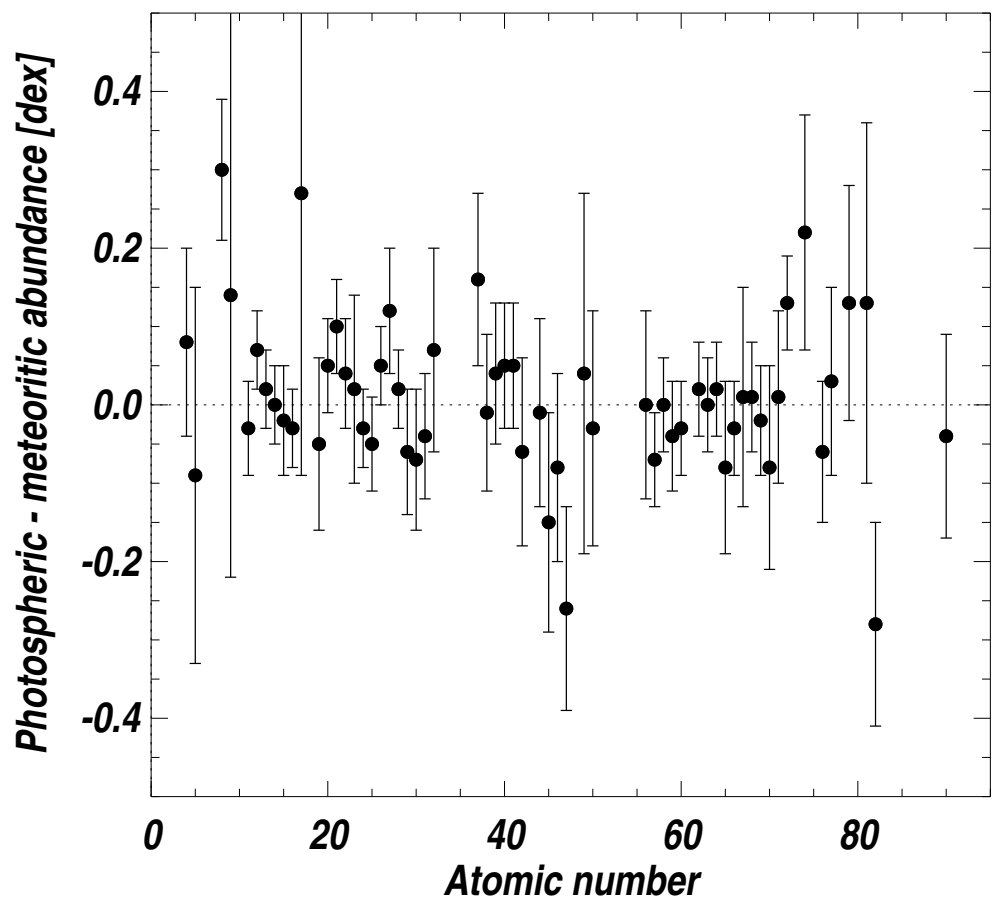,width=15cm}}
\caption{Difference between the logarithmic abundances determined from the 
solar photosphere and the CI carbonaceous chondrites as a function of atomic number.
With a few exceptions the agreement is excellent. Note that due to depletion in the
Sun and meteorites, the data points for Li, C, N and the noble gases fall outside the range of the figure. 
\label{f:sunvsmet}
}
\end{figure}

The mean abundances of CI chondrites presented in Table \ref{t:sun} have been 
appropriated from the data carefully compiled by 
\citet{2009arXiv0901.1149L}. 
To place the meteoritic abundances (defined as $N_{\rm Si} = 10^6$) 
on the same absolute scale as for the photosphere (defined as $\log \epsilon_{\rm H} = 12.00$),
we have renormalized their data 
such that the meteoritic and photospheric Si abundances agree.
The relation between the two scales is thus specified as
$\log \epsilon_{\rm X} = 1.51 + \log N_{\rm X}$.
In principle the conversion factor could employ additional elements besides Si; 
\citet{1989GeCoA..53..197A} 
for example used Na, Mg, Si, Ca, V, Cr, Co, Ni, Y, Zr, Nb and Mo, while
\citet{2009arXiv0901.1149L} 
included as many as 39 elements.
In practice, however, this does not change the result noticeably but it introduces
some degree of arbitrariness in the selection of elements to include in this exercise.
Fig. \ref{f:sunvsmet} shows the difference between the thus obtained 
photospheric and CI chondrite abundances, highlighting the excellent
correspondence between the two methods. 
Naturally all volatile elements are depleted in the meteorites,
whereas  Li is depleted in the solar photosphere.
For the other 45 elements with a purported photospheric uncertainty less than 25\%,
the mean difference between photospheric and meteoritic abundances is $0.00 \pm 0.05$\,dex. 
The photospheric and meteoritic abundances 
differ by more than the total combined uncertainty for only 10 out of the 57 non-volatile elements
(expected $\approx18/57$), 
which suggests that our estimated errors are in fact slightly too conservative.
It should be emphasized that this excellent agreement cannot be interpreted
as an absence of any significant diffusion in the Sun (Sect. \ref{s:bulk}), as the meteoritic scale is set by Si;
all elements heavier than He are fractionated by $\approx 0.04$\,dex relative to H but hardly at all
relative to each other. 

In spite of the excellent agreement overall illustrated in Fig. \ref{f:sunvsmet}, 
there are a few noticeable exceptions.
Of the ten non-depleted elements where the solar and meteoritic abundances differ
by more than 0.1\,dex (25\%), this can in all cases be blamed on uncertain
photospheric analyses. 
Indeed, in many cases the two methods still agree marginally within the combined errors
(Cl, Au, Tl).
We note a significant improvement for W 
compared with previous compilations of the solar chemical composition 
($\Delta \log \epsilon = +0.49$\,dex in 
\citet{2005ASPC..336...25A}
while here the difference amount to $+0.20$\,dex);
we suspect that unidentified blends and/or erroneous continuum placement 
may be at fault for the persisting discrepancy, as well as for Rb and Hf.
For Co we reckon that the non-LTE corrections by 
\citet{2008PhST..133a4013B} 
are somewhat overestimated.
The remaining three elements Rh, Ag and Pb have had their photospheric
values determined from transitions originating either in the ground state
or from low-excitation levels of the neutral species (also true for Pd, W and Au), 
which are prone to non-LTE effects in their excitation and ionization populations
\citep{2005ARA&A..43..481A}. 

\subsection{Solar Neighborhood}
\label{s:solarneighborhood}


Independent estimates of the chemical composition of the Sun are provided by
solar-type stars, OB stars, H\,{\sc ii} regions like
the Orion nebula and the interstellar medium in the solar neighborhood.
In order to make a meaningful comparison one must first account for the effects of
diffusion in the Sun and Galactic chemical enrichment. The former amounts to $0.04$\,dex for metals
\citep[][ cf.~Sect.~\protect\ref{s:bulk}]{2002JGRA..107.1442T} 
while the latter is typically predicted to be $0.05-0.15$\,dex depending on the element in question
\citep{2003MNRAS.339...63C, 2008EAS....32..311P}. 

Whether the Sun is somehow special in its chemical composition relative to
other FG dwarfs in the Galactic thin disk has attracted attention over the years
\citep[e.g.][ and references therein]{2008PhST..130a4036G, 2008ApJ...684..691R}. 
It should be borne in mind that most large-scale abundance analyses of Galactic thin disk stars to date
\citep[e.g.][]{
1993A&A...275..101E, 
2003MNRAS.340..304R, 
2005A&A...433..185B, 
2007A&A...465..271R, 
2008MNRAS.384..173F} 
have been performed differentially relative to the Sun to remove the uncertainties from $gf$-values
and to minimize systematic errors due to inadequacies in the stellar modelling. 
These studies therefore reveal little about the absolute abundances of the stars
but do imply that the Sun is a rather ordinary thin disk G dwarf for its age.
In his volume-complete sample of stars within 25\,pc, 
\citet{2008MNRAS.384..173F} 
found a mean spectroscopic metallicity of \feh $=-0.02 \pm 0.18$. 
Similarly, among the stars within 40\,kpc having an isochrone age of 4-6\,Gyr,
\citet{2008arXiv0811.3982H} 
obtained a mean metallicity of \feh $=-0.00 \pm 0.10$.
%
In their study of 11 solar twins, 
\citet{melendez_twins} 
found that the solar X/Fe ratios did not differ by more than $0.06$\,dex from the mean
of the solar twin sample for any of the 23 investigated elements. 

In the past, the Sun appeared to be metal-rich
compared with, for example, young OB stars in the solar neighborhood, 
despite the expected chemical enrichment
over the past 4.5\,Gyr.
In the meantime, the solar abundances have decreased and the hot star
abundances have increased. The latter have also become more accurate, as reflected in greatly diminished
uncertainties
\citep[e.g.][]{
2004ApJ...617.1115D, 
2006ApJ...639L..39N, 
2008arXiv0811.4114M}. 
Arguably the most accurate result to date is the analysis of 
\citet{2008ApJ...688L.103P}, 
who have carried out a non-LTE study of six nearby B-type main sequence stars.
Their results are
presented in Table \ref{t:neighborhood}; for S we give the mean abundance from 
\citet{2006A&A...457..651M} 
while the Ar abundance is taken from 
\citet{2008ApJ...678.1342L}. 
When accounting for diffusion in the Sun, there is good agreement between
the solar and hot star abundances for O, Ne and Mg in particular. While the situation 
has greatly improved compared with the solar abundances of 
\citet{1998SSRv...85..161G} 
and earlier compilations, there is still a tendency for the solar C, N and Fe abundances to
be somewhat higher than in the B stars, contrary to expectations. 
It is unclear whether the solution can be found in the solar or B star analyses or, 
if a real difference indeed exists, perhaps due to infall of low-metallicity gas to the solar neighborhood. 

The abundances of nearby H\,{\sc ii} regions like the Orion nebula also provide
constraints on the solar chemical composition. 
In Table \ref{t:neighborhood} we provide the values from
\citet{2004MNRAS.355..229E}, 
\citet{2005ApJ...618L..95E} 
and
\citet{ 2007ApJ...670..457G} 
interpolated to a Galacto-centric radius of 8\,kpc.
The observed scatter between individual H\,{\sc ii} regions is now quite small
and the inferred present-day Galactic abundance gradient well determined. 
It is worth remembering that there are multiple potential pitfalls
in the analyses of H\,{\sc ii} regions, including corrections for unseen
ionization stages and the effects of temperature fluctuations in the nebula. 
Furthermore, abundances derived from recombination (permitted) lines are often
significantly larger than the corresponding values from collisionally excited (forbidden) lines.
Finally, dust condensation needs to be taken into account for some elements. 
For C and O this somewhat uncertain dust correction amounts to $\approx 0.1$\,dex in Orion
\citep{2004MNRAS.355..229E}. 
Volatile elements like Ne and S are not expected to form dust
but refractory elements like Mg and Fe are predominantly in the solid phase, so it is
therefore not possible to deduce a meaningful total elemental abundance for them. 
Overall, the agreement between the Sun and H\,{\sc ii} regions is very satisfactory
when accounting for diffusion and a minor chemical enrichment over the past 4.5\,Gyr. 

Because of the generally lower temperatures encountered 
and the consequentially larger degree of dust condensation, the interstellar medium
is a less reliable chemical composition gauge than H\,{\sc ii} regions
\citep{2004oee..symp..336J}. 
Indeed, rather than setting constraints on the solar abundances, the Sun is often used
to infer the amount in the solid phase in the interstellar medium
\citep[e.g.][]{2001ApJ...554L.221S}. 
The local abundances of the volatile elements C
\citep{2004ApJ...605..272S}, 
N 
\citep{2007ApJ...654..955J} 
and O 
\citep{
2006ApJ...641..327C, 
2005ApJ...619..891J} 
in the interstellar medium support our recommended solar values in general.
With the older, high ($\log \epsilon_{\rm O} = 8.83-8.93$)
solar abundances preferred by
\citet{1998SSRv...85..161G} 
and
\citet{1989GeCoA..53..197A} 
it becomes very challenging to hide sufficient amounts of O in dust
\citep{2007PhDT.........3J}. 
The gas phase abundances of the refractory elements is only a small fraction of the total
abundance in both the cold and warm interstellar medium, 
preventing them from setting any stringent constraints.


Also listed in Table \ref{t:neighborhood} are predictions from models of
Galactic chemical evolution 
\citep{2003MNRAS.339...63C}. 
Indeed, with these values it becomes difficult to reconcile the solar abundances
with the results from OB stars and H\,{\sc ii} regions for some elements like Fe.
It is important to bear in mind though that the models have varying degrees of 
reliability for different elements. 
The evolutionary predictions for O, Ne and Mg should be the most accurate,
as they are produced in hydrostatic burning and ejected by SNe\,II; in fact
for these elements there is good agreement between the Sun and its local neighborhood.
All other elements are either produced in explosive burning and therefore dependent on
the adopted mass-cut in the supernova explosion, or have a non-negligible contribution
from SNe\,Ia or AGB stars, making theoretical estimates rather uncertain. 
One may conclude that the local Galactic volume has had a smaller star formation rate
than assumed in these chemical evolution models. 
This inference is born out empirically by the observed age-metallicity relation, which 
\citet{2008arXiv0811.3982H} 
estimated to be only $0.018$\,dex/Gyr for \feh\ in the solar neighborhood, 
i.e. only half of the predictions given in Table \ref{t:neighborhood}.
An alternative explanation would be recent Galactic infall of more pristine gas.


\subsection{Helioseismology}
\label{s:helioseismology}

\begin{figure}
\centerline{\psfig{figure=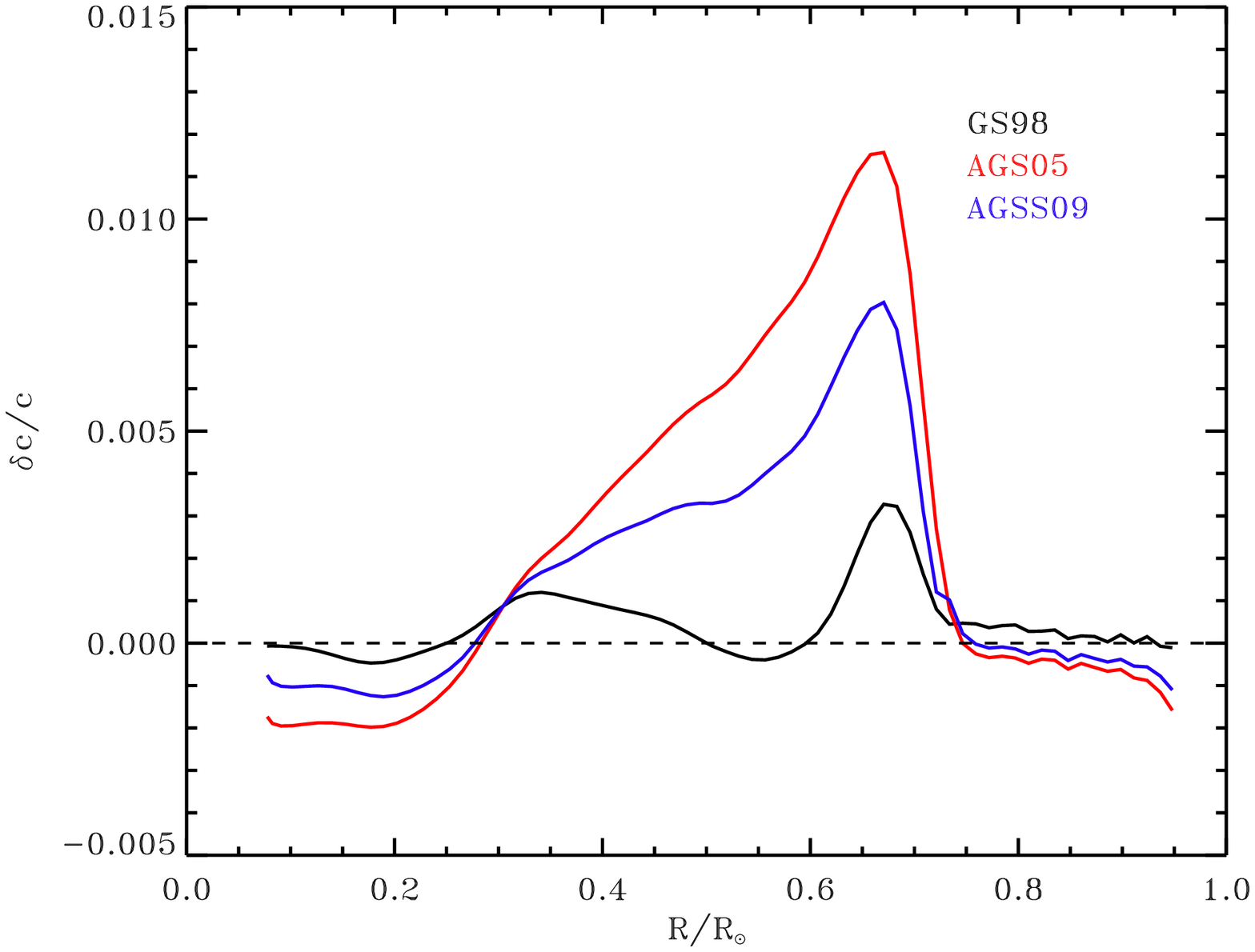,width=13cm}}
\caption{The differences between the helioseismic and
predicted sound speeds as a function of depth
\citep{serenelli_helio}. 
The standard solar models shown here only differ in the assumed chemical compositions:
\citet{1998SSRv...85..161G} 
(black line, here denoted GS98),
\citet{2005ASPC..336...25A} 
(red line, AGS05)
and the present work
(blue line, AGSS09).
Each model has independently been calibrated to
achieve the correct solar luminosity, temperature and age.  
The base of the convection zone is at $R = 0.71$\,R$_\odot$, which
is also where the discrepancy starts in earnest
in all three cases.
\label{f:soundspeed}
}
\end{figure}

The situation is decidedly less encouraging when attempting to assess the solar abundances
by means of helioseismology.
Since the various solar oscillation p-modes penetrate to different depths, it is possible
to thereby map the variation of the sound speed in the solar interior, which can be compared with
the predicted values from solar structure models
\citep[see][ for a detailed review]{2008PhR...457..217B}. 
With the
\citet{1998SSRv...85..161G} 
abundances, the predicted sound speed as a function of depth from standard solar models shows a
very good agreement with the helioseismic values.
When instead employing the revised values of 
\citet{2005ASPC..336...25A}, 
this is no longer the case 
\citep[e.g.][]{
2004ApJ...606L..85B, 
2004PhRvL..93u1102T, 
2005ApJ...618.1049B, 
2005ApJ...627.1049G, 
2006ApJ...649..529D, 
2008PhR...457..217B}. 
In this context it is perhaps fair to say that `better is worse'. 
The sound speed discrepancy most clearly shows up immediately
below the bottom of the convection zone, as illustrated in Fig. \ref{f:soundspeed}.
Also with the 
\citet{1998SSRv...85..161G} 
abundances there was a disagreement in this region but it is greatly aggravated by 
the further lowering of the solar O and Ne abundances and to a lesser extent by the reassessment of C, N and Fe.
With the solar chemical composition recommended here, the situation is alleviated somewhat due
to the slightly higher abundances of these elements and consequently larger opacity in the radiative interior
but the deviation is still highly significant.

It is not only the sound speed variation that is at variance when using the new solar abundances.
The inferred depth of the convection zone is now too shallow: $R_{\rm BCZ} \approx 0.725$\,R$_\odot$
instead of the helioseismic measurement $0.7133 \pm 0.0005$\,R$_\odot$ 
\citep{2004ApJ...606L..85B}. 
Furthermore, the resulting He abundance when calibrating
solar interior models to achieve the correct solar luminosity and temperature at the solar age
is similarly inconsistent with the value inferred from helioseismology (see Sect. \ref{s:noble}):
$Y_{\rm S} \approx 0.238$ compared with the observed 
$0.2485 \pm 0.0034$
\citep{2004ApJ...606L..85B}, 
where $Y_{\rm S}$ is the present-day He mass fraction at the surface.
The problem also persists in the solar core both as implied by low-degree p-modes
\citep{2007ApJ...670..872C} 
and by gravity modes
\citep{2007Sci...316.1591G}. 
Thus the problem exists both in the convective envelope, the radiative interior and in the core.

Following the initial revisions of the solar C and O abundances by
\citet{2001ApJ...556L..63A} 
and
\citet{2002ApJ...573L.137A}, 
many possible solutions have been put forward to explain the problem
albeit to date with little success. 
The most obvious explanation would be that the opacities are underestimated.
\citet{2005ApJ...618.1049B} 
estimated that a $10-20$\% increase of the OPAL opacities over a relatively substantial
fraction of the solar interior ($R = 0.4-0.7$\,R$_\odot$ or equivalently $T = 2-5\cdot 10^6$\,K)
would be required;
\citet{2009A&A...494..205C} 
found even slightly more stringent requirements with the 
\citet{2005ASPC..336...25A} 
solar abundances. 
\citet{serenelli_helio} 
have investigated the differences with the solar chemical composition presented here
and conclude that $\le 12$\% extra opacity is required to restore the previous agreement in sound speed.
Little suggests, however, that current atomic physics calculations for stellar interiors are missing such
a substantial fraction of the opacity.
For example, \citet{2005MNRAS.360..458B} 
showed that opacities from the Opacity Project are at most $3$\% larger than the corresponding OPAL
values in the relevant temperature range. 
While unlikely to cure the problem by itself, it is perhaps wise to bear in mind that a 
similar situation for pulsating stars in fact existed in the 1980s, 
which found an unexpected resolution with the more comprehensive calculations by 
the OPAL and OP collaborations.

There are other ways to compensate for the lower opacities resulting from the 
lower C, N, O and Ne abundances. 
One is that the radiative interior contains more metals than expected from the photospheric
abundances and standard models for diffusion. 
\citet{2004A&A...417..751A} 
first suggested that the diffusion efficiency may be seriously underestimated
but the required elemental settlement is about twice the predicted value.
Furthermore, while this may remedy the sound speed discrepancy, problems still
remain in terms of the He abundance and depth of the convection zone
\citep{
2004ESASP.559..574M, 
2005ApJ...627.1049G, 
2007ApJ...658L..67Y}. 
A hypothetical late accretion of metal-depleted gas during the formation of the Sun
runs into the same trouble
\citep{2005ApJ...627.1049G, 2007A&A...463..755C}. 
Alternatively, the reduced O abundance could conceivably be compensated by a corresponding but larger increase
in the Ne abundance
\citep{2005ApJ...620L.129A}. 
\citet{2005ApJ...631.1281B} 
estimated that a factor of three ($0.5$\,dex) higher Ne content compared with the recommended values by
\citet{2005ASPC..336...25A}, 
together with minor adjustments of C, N, O and Ar within their associated uncertainties would restore
the  good agreement found with the 
\citet{1998SSRv...85..161G} 
solar composition. There was initial excitement when 
\citet{2005Natur.436..525D} 
argued for exactly such an increase in Ne based on inferences from coronal spectroscopy of other stars
but there is in fact little support for this idea, as explained in Sect. \ref{s:noble}. 
The here recommended abundance of Ne as well as of C, N, O and Fe are $\approx 0.05$\,dex
higher than in 
\citet{2005ASPC..336...25A}. 
This shift only partly allays the solar modelling discrepancy, as seen in Fig. \ref{f:soundspeed}. 

Given that the problem exceeds the likely uncertainties in existing input data (e.g. opacities, diffusion, abundances),
the only remaining option on the solar interior side seems to be to invoke some physical process not
yet accounted for in standard solar models. 
It is perhaps suggestive that the largest deviation occurs immediately below the convection zone.
\citet{2005ASPC..336..235A} 
have proposed that internal gravity waves generated by the convective motion at the base of the convection zone
propagate inwards and deposit their energy in the radiative zone.
The ensuing structure changes are similar to an opacity enhancement, thus going in the right direction. 
In addition, the gravity waves induce additional mixing, which would also help.
\citet{2005ApJ...618..908Y} 
have investigated these processes using multi-dimensional hydrodynamical simulations of stellar convection zones
and found qualitative improvements in many stellar regimes. 
Due to the computationally very challenging nature of the hydrodynamics, 
the modelling is unfortunately not yet able to quantitatively predict whether the effects are sufficient
to explain the helioseismic signature. 

Thus, most suggestions to resolve the solar modelling discrepancy 
have already been ruled out or are considered improbable.
The few possibilities still standing do so largely because no detailed modelling 
has as yet been forthcoming to properly put the proposals to test. 
It is possible though that a combination of several factors could be the explanation, but
such fine-tuning would seem rather contrived.
Should no likely solution be found, it would suggest 
that the photospheric abundances presented herein are wrong. 
Given the discussion in Sect. \ref{s:ingredients} and Sect. \ref{s:solarabund}, it is not easy
to understand where such a serious error may lie but the possibility should obviously not 
be discounted.

\subsection{Solar neutrinos}
\label{s:neutrinos}


Neutrinos provide another means by which to infer the solar chemical composition,
though current experiments are not yet sufficiently sensitive to return decisive results in this respect. 
The metallicity adopted in a solar model modifies the conditions in the core,
and thus the neutrino fluxes produced by the pp-chain and CNO-cycle. 
Whilst the neutrino fluxes cannot be directly associated with abundances of
individual elements, the relevant neutrino channels depend most sensitively on C, N, O, Si and Fe
\citep{2005ApJ...626..530B}. 
Using neutrino data from the SuperKamiokande I+II, SNO and Borexino experiments,
\citet{2008arXiv0811.2424P} 
have derived the neutrino oscillations and fluxes expected from solar models constructed with the 
\citet{1998SSRv...85..161G} 
and
\citet{2005ASPC..336...25A} 
compositions.
Intriguingly, the existing data imply $^7$Be and $^8$B neutrino fluxes 
that fall between the predictions for the two models;
we suspect that the solar neutrinos would favour the new abundances presented herein.
A more direct test of the heavily debated solar C, N and O abundances would come from
measuring the $^{13}$N and $^{15}$O $\beta$-decay neutrinos that follow proton capture on
$^{12}$C and $^{14}$N, respectively.
Meaningful constraints should come from Borexino within the next couple of years, and
the proposed SNO+ experiment should weigh in with a definite answer sometime after 2011
\citep{2008ApJ...687..678H}. 

\section{CONCLUDING REMARKS}
\label{s:conclusions}

\subsection{Summary Points}
\label{s:summary}


We have critically examined all ingredients for the determination of the solar photospheric 
chemical composition, from the atmospheric
and line formation modelling, required atomic data and selection of
lines to the final estimate of the elemental abundances and associated errors
(Sect. \ref{s:solarabund}). 
A notable feature of this analysis is that 
for essentially every element, the effects of 3D hydrodynamical model atmospheres
have now been accounted for.
The 3D solar model employed here outperforms any 1D model, theoretical or semi-empirical,
in a raft of observational tests. 
The values we recommend for the most abundant metals C, N, O, Ne and Fe are
significantly smaller (by $\sim 0.2$\,dex) than those advocated in the widely-used compilation of
\citet{1989GeCoA..53..197A}. 
Most of these differences can be attributed to the improvements over the past
two decades in modelling the solar atmosphere, allowing for
departures from LTE in the line formation, properly treating blends and
employing improved atomic and molecular data (Sect. \ref{s:ingredients}). 
Amongst other things, these new, lower abundances are supported by the excellent
agreement between different abundance indicators for C, N and O: 
low-excitation forbidden atomic lines, high-excitation permitted atomic lines and
various molecular features. 

A comparison of the photospheric abundances and the most primitive 
class of meteorites, the CI chondrites, reveals a very good agreement for 
nearly all non-volatile elements (Sect. \ref{s:meteorites}). 
Since the volatile elements, including H, have been
depleted in the chondrites, the meteoritic abundance scale has been placed
on the same absolute level as the photospheric abundances by enforcing
equivalence of the Si abundances. The remaining few elements (Rb, Rh, Pd, Ag, Hf, Pb) that
suggest a significant difference between meteorites and the Sun can probably 
be attributed to poor solar determinations, likely due to incorrect 
$gf$-values, unidentified blends or unaccounted non-LTE effects. 

In order to make a meaningful comparison between the Sun and the abundances
measured in the solar neighborhood, one must first correct for the effects
of diffusion in the Sun and chemical enrichment in the Galaxy over the past 4.56\,Gyr. 
All elements are predicted to have settled to the solar interior by essentially the same amount, meaning
that the bulk, or equivalently the proto-solar, composition is higher by $0.04$\,dex 
than listed in Table \ref{t:sun} for all elements heavier than He  (Sect. \ref{s:bulk}). 
Our recommended solar abundances agree very well in general 
with those measured in nearby OB stars, H\,{\sc ii} regions like Orion and the 
local ISM  (Sect. \ref{s:solarneighborhood}). 
We note that care should be exercised when comparing our solar results with corresponding
abundances for late-type stars derived with classical 1D model atmospheres;
ideally a differential study of the stars and the Sun should be performed and using
the solar abundances presented in Table \ref{t:sun} to place them on an absolute scale.

In spite of the many apparent successes of the solar 
chemical composition we recommend here,
the outlook is not entirely rosy.
With the downward revision of the overall metal content presented here compared to the
\citet{1998SSRv...85..161G} 
abundances, the previous excellent agreement between predicted sound speeds from
standard solar models and those inferred from helioseismology is ruined 
(Sect. \ref{s:helioseismology}).
To date there has been no fully convincing solution put forward.
The discordance has been alleviated somewhat relative to the recommended values in 
\citet{2005ASPC..336...25A} 
but it nevertheless remains a significant discrepancy in urgent need of resolution.

\subsection{Future Issues}
\label{s:future}


While major progress has been made in recent years in
the inferred solar chemical composition, there is no question that a great deal of
work remains before we can confidently state that
each solar elemental abundance is known to better than 0.01\,dex 
($2\%$). 
For many, if not most elements, the achieved accuracy today is 
more like 0.05\,dex or indeed significantly worse in many cases. 
A concerted effort on multiple fronts is required to substantially decrease the remaining uncertainties.

We do not consider it likely that
subsequent generations of 3D hydrodynamical model atmospheres will dramatically (i.e. at the $>0.1$\,dex
level) amend the inferred solar photospheric chemical composition, given how successful
the most recent models are in reproducing key observational diagnostics.
It is fair to say that the mean temperature structure of the quiet solar granulation,
the velocity distribution and the atmospheric inhomogeneities
are now rather well known, at least in the formation regions of the weaker lines largely
employed by the abundance studies described herein. 
Higher atmospheric layers, to which the strongest atomic and molecular lines are more sensitive, 
are naturally less constrained observationally; on the modelling side this
requires improved radiative transfer treatment and allowance of departures from LTE for
key elements. 
For completeness, one should include the effects of magnetic fields in the 3D solar atmospheric
models, but we do not expect that these will noticeably influence the derived abundances
\citep{2009LRSP....6....2N}. 

The greatest uncertainties afflicting solar abundance work today instead stem from 
poor atomic and molecular data, and incomplete modelling of departures from LTE. 
Few spectral lines have their transition probabilities determined to better than a few \%. 
As an example here it suffices to mention the case of Mg with its relatively simple atomic structure:
the majority of the Mg\,{\sc i} lines employed here and elsewhere for solar abundance purposes still have
quoted uncertainties of $25-50$\% according to the NIST database, which is clearly unacceptable. 
Fortunately there has been some laudable progress on this score in recent years, both in terms of
experimental efforts
\citep[e.g.][]{2003ApJ...584L.107J, 2009ApJS..182...51L, 2007A&A...472L..43B},
and theoretical calculations 
\citep[e.g.][]{1992RMxAA..23...45K, 1998A&A...337..495P, 2005MNRAS.360..458B}
as discussed in Sect. \ref{s:inputdata}.
A major problem, however, is the dearth of atomic physicists devoted to satisfying the 
astronomical need for various types of input data; 
the astronomical community ought to better appreciate and encourage the relatively
few individuals carrying out this important work, given its tremendous benefit for 
solar and stellar astrophysics especially. 
On the wish-list are improved $gf$-values for \mgi, \mgii, \siii, \caii, 
the once-ionized species of the Fe-peak elements,  \cui\ and \srii.
Also, atomic lines in the IR have great potential given the accurate determination
of the continuum and few blends but more often than not accurate transition probabilities are lacking.

Equally important is the consideration of non-LTE effects on line formation.
As discussed by \citet{2005ARA&A..43..481A}, 
even though departures from LTE are expected to be relatively serene in the Sun, 
one should always expect non-LTE abundance corrections at around the 0.05\,dex level;
indeed in many cases the effects will be significantly more severe. 
Still today, most elements have not been exposed to a detailed non-LTE study even in the Sun.
Even rarer are solar abundance works going beyond the 1D framework and employing
3D non-LTE calculations 
\citep[e.g.][]{
2003A&A...399L..31A, 
2004A&A...417..751A}. 
In part, the scarcity of non-LTE studies is related to a lack of
adequate atomic data. 
Compared with the LTE case, vastly more input data are required in non-LTE calculations
than are presently available for most elements. 
Significant headway will only become possible with the advent of detailed photo-ionization and
collisional cross-sections; for the latter, estimates of the impact of both electron and hydrogen collisions, as well as 
other processes like charge transfer, are urgently needed.  
There is also an issue of manpower; 
careful non-LTE studies are very time-consuming and there
are not enough specialists working in the field to satisfy the demand.
Finally, we note that non-LTE line formation should be explored in 3D model atmospheres, 
since the the presence of atmospheric inhomogeneities 
is generally expected to amplify any existing non-LTE effects seen in 1D.

A section on future issues in the solar chemical composition 
would not be complete without stressing the importance of finally resolving
the solar modelling problem uncovered by helioseismology (Sect. \ref{s:helioseismology}). 
Many hypotheses have been put forward to explain the discrepancy, 
but no clear solution has been forthcoming. 
Until an explanation has been found, a mutual cloud of doubt will always linger
over the solar chemical composition recommended here and current models of the solar interior.
Whether the problem will be traced in the end to shortcomings in the atmospheric or interior
modelling remains to be seen, but whatever the resolution it will place solar and stellar
astrophysics -- and by extension most of astronomy -- on a much firmer footing.

\subsection{Disclosure Statement}

The authors are not aware of any affiliations, memberships, funding, or
financial holdings that might be perceived as affecting
the objectivity of this review.

\subsection{Acknowledgements}


We are indebted to our collaborators over the years
for their important contribution to uncovering the solar chemical composition.  We would particularly like to highlight the contributions of
Carlos Allende Prieto, David Lambert, Jorge Mel{\'e}ndez and Tiago Pereira.
The work presented here would not have been possible without 
extensive efforts in solar atmospheric and line-formation modelling by our colleagues
Mats Carlsson, Remo Collet, 
Wolfgang Hayek, 
Dan Kiselman, \AA ke Nordlund, Bob Stein and Regner Trampedach.
We thank Tiago Pereira and Aldo Serenelli for providing figures. 
Finally, we acknowledge a large number of 
colleagues around the world for many stimulating discussions, healthy and friendly competition, 
providing information prior to publication and invaluable comments on the manuscript
and topics discussed herein.



\bibliography{sun} 


\newpage

\begin{table}[!t]
\caption{Element abundances in the present-day solar photosphere. Also given are the corresponding
values for CI  carbonaceous chondrites 
\citep{2009arXiv0901.1149L}. 
Indirect photospheric estimates have been used for the noble gases (Sect. \ref{s:noble}).
\label{t:sun}}
\smallskip
\begin{tabular}{llcc|llcc}
\hline
 \noalign{\smallskip}
 & Elem.  &  Photosphere  &   Meteorites &
 & Elem.  &  Photosphere  &   Meteorites  \\
 \noalign{\smallskip}
 \hline
 \noalign{\smallskip}
1  & H   & $12.00$            &  $8.22 \pm 0.04$        & 44 & Ru  &  $1.75 \pm 0.08$   &  $1.76 \pm 0.03$  \\
2  & He  & $[10.93 \pm 0.01]$ &  $1.29$                 & 45 & Rh  &  $0.91 \pm 0.10$   &  $1.06 \pm 0.04$  \\
3  & Li  &  $1.05 \pm 0.10$   &  $3.26 \pm 0.05$        & 46 & Pd  &  $1.57 \pm 0.10$   &  $1.65 \pm 0.02$  \\
4  & Be  &  $1.38 \pm 0.09$   &  $1.30 \pm 0.03$        & 47 & Ag  &  $0.94 \pm 0.10$   &  $1.20 \pm 0.02$  \\
5  & B   &  $2.70 \pm 0.20$   &  $2.79 \pm 0.04$        & 48 & Cd  &                    &  $1.71 \pm 0.03$  \\
6  & C   &  $8.43 \pm 0.05$   &  $7.39 \pm 0.04$        & 49 & In  &  $0.80 \pm 0.20$   &  $0.76 \pm 0.03$  \\
7  & N   &  $7.83 \pm 0.05$   &  $6.26 \pm 0.06$        & 50 & Sn  &  $2.04 \pm 0.10$   &  $2.07 \pm 0.06$  \\
8  & O   &  $8.69 \pm 0.05$   &  $8.40 \pm 0.04$        & 51 & Sb  &  $$                &  $1.01 \pm 0.06$  \\
9  & F   &  $4.56 \pm 0.30$   &  $4.42 \pm 0.06$        & 52 & Te  &                    &  $2.18 \pm 0.03$  \\
10 & Ne  &  $[7.93 \pm 0.10]$ &  $-1.12$                & 53 & I   &                    &  $1.55 \pm 0.08$  \\
11 & Na  &  $6.24 \pm 0.04$   &  $6.27 \pm 0.02$        & 54 & Xe  &  $[2.24 \pm 0.06]$ &  $-1.95$  \\
12 & Mg  &  $7.60 \pm 0.04$   &  $7.53 \pm 0.01$        & 55 & Cs  &                    &  $1.08 \pm 0.02$  \\
13 & Al  &  $6.45 \pm 0.03$   &  $6.43 \pm 0.01$        & 56 & Ba  &  $2.18 \pm 0.09$   &  $2.18 \pm 0.03$  \\
14 & Si  &  $7.51 \pm 0.03$   &  $7.51 \pm 0.01$        & 57 & La  &  $1.10 \pm 0.04$   &  $1.17 \pm 0.02$  \\
15 & P   &  $5.41 \pm 0.03$   &  $5.43 \pm 0.04$        & 58 & Ce  &  $1.58 \pm 0.04$   &  $1.58 \pm 0.02$  \\
16 & S   &  $7.12 \pm 0.03$   &  $7.15 \pm 0.02$        & 59 & Pr  &  $0.72 \pm 0.04$   &  $0.76 \pm 0.03$  \\
17 & Cl  &  $5.50 \pm 0.30$   &  $5.23 \pm 0.06$        & 60 & Nd  &  $1.42 \pm 0.04$   &  $1.45 \pm 0.02$  \\
18 & Ar  &  $[6.40 \pm 0.13]$   &  $-0.50$              & 62 & Sm  &  $0.96 \pm 0.04$   &  $0.94 \pm 0.02$  \\
19 & K   &  $5.03 \pm 0.09$   &  $5.08 \pm 0.02$        & 63 & Eu  &  $0.52 \pm 0.04$   &  $0.51 \pm 0.02$  \\
20 & Ca  &  $6.34 \pm 0.04$   &  $6.29 \pm 0.02$        & 64 & Gd  &  $1.07 \pm 0.04$   &  $1.05 \pm 0.02$  \\
21 & Sc  &  $3.15 \pm 0.04$   &  $3.05 \pm 0.02$        & 65 & Tb  &  $0.30 \pm 0.10$   &  $0.32 \pm 0.03$  \\ 
22 & Ti  &  $4.95 \pm 0.05$   &  $4.91 \pm 0.03$        & 66 & Dy  &  $1.10 \pm 0.04$   &  $1.13 \pm 0.02$  \\
23 & V   &  $3.93 \pm 0.08$   &  $3.96 \pm 0.02$        & 67 & Ho  &  $0.48 \pm 0.11$   &  $0.47 \pm 0.03$  \\
24 & Cr  &  $5.64 \pm 0.04$   &  $5.64 \pm 0.01$        & 68 & Er  &  $0.92 \pm 0.05$   &  $0.92 \pm 0.02$  \\
25 & Mn  &  $5.43 \pm 0.05$   &  $5.48 \pm 0.01$        & 69 & Tm  &  $0.10 \pm 0.04$   &  $0.12 \pm 0.03$  \\
26 & Fe  &  $7.50 \pm 0.04$   &  $7.45 \pm 0.01$        & 70 & Yb  &  $0.84 \pm 0.11$   &  $0.92 \pm 0.02$  \\
27 & Co  &  $4.99 \pm 0.07$   &  $4.87 \pm 0.01$        & 71 & Lu  &  $0.10 \pm 0.09$   &  $0.09 \pm 0.02$  \\
28 & Ni  &  $6.22 \pm 0.04$   &  $6.20 \pm 0.01$        & 72 & Hf  &  $0.85 \pm 0.04$   &  $0.71 \pm 0.02$  \\
29 & Cu  &  $4.19 \pm 0.04$   &  $4.25 \pm 0.04$        & 73 & Ta  &                    &  -$0.12 \pm 0.04$ \\
30 & Zn  &  $4.56 \pm 0.05$   &  $4.63 \pm 0.04$        & 74 & W   &  $0.85 \pm 0.12$   &  $0.65 \pm 0.04$  \\
31 & Ga  &  $3.04 \pm 0.09$   &  $3.08 \pm 0.02$        & 75 & Re  &                    &  $0.26 \pm 0.04$  \\
32 & Ge  &  $3.65 \pm 0.10$   &  $3.58 \pm 0.04$        & 76 & Os  &  $1.40 \pm 0.08$   &  $1.35 \pm 0.03$  \\
33 & As  &                    &  $2.30 \pm 0.04$        & 77 & Ir  &  $1.38 \pm 0.07$   &  $1.32 \pm 0.02$  \\
34 & Se  &                    &  $3.34 \pm 0.03$        & 78 & Pt  &                    &  $1.62 \pm 0.03$  \\
35 & Br  &                    &  $2.54 \pm 0.06$        & 79 & Au  &  $0.92 \pm 0.10$   &  $0.80 \pm 0.04$  \\
36 & Kr  &  $[3.25 \pm 0.06]$   &  $-2.27$              & 80 & Hg  &                    &  $1.17 \pm 0.08$  \\
37 & Rb  &  $2.52 \pm 0.10$   &  $2.36 \pm 0.03$        & 81 & Tl  &  $0.90 \pm 0.20$   &  $0.77 \pm 0.03$  \\
38 & Sr  &  $2.87 \pm 0.07$   &  $2.88 \pm 0.03$        & 82 & Pb  &  $1.75 \pm 0.10$   &  $2.04 \pm 0.03$  \\
39 & Y   &  $2.21 \pm 0.05$   &  $2.17 \pm 0.04$        & 83 & Bi  &  $$                &  $0.65 \pm 0.04$  \\
40 & Zr  &  $2.58 \pm 0.04$   &  $2.53 \pm 0.04$        & 90 & Th  &  $0.02 \pm 0.10$   &  $0.06 \pm 0.03$  \\
41 & Nb  &  $1.46 \pm 0.04$   &  $1.41 \pm 0.04$        & 92 & U   &                    &  -$0.54 \pm 0.03$  \\
42 & Mo  &  $1.88 \pm 0.08$   &  $1.94 \pm 0.04$        & $$\\
\noalign{\smallskip}
\hline
\end{tabular}
\end{table}

\newpage

\begin{table}[!t]
\caption{The derived photospheric solar C, N and O abundances as 
indicated by various atomic and molecular species and a 3D hydrodynamical
model of the solar atmosphere
\citep{sun_c, sun_n, sun_o}. 
The corresponding results obtained with the temporally and spatially (over equal optical depth
surfaces) averaged 3D model (denoted $<3$D$>$),
the 1D theoretical {\sc marcs} model
\citep{2008A&A...486..951G} 
and the 1D semi-empirical model of 
\citet{1974SoPh...39...19H} 
are also given for comparison purposes. The errors given here are only the
standard deviation ($\sigma$) for each type of transition. For the atomic transitions, departures
from LTE have been considered. 
\label{t:cno}
}
\smallskip
\begin{center}
\begin{tabular}{lcccc}
\toprule
\noalign{\smallskip}
lines  & \multicolumn{4}{c}{${\rm log} \epsilon_{\rm C, N, O}$} \\
\noalign{\smallskip}
\cline{2-5}
\noalign{\smallskip}
        &   3D & $<3$D$>$ & HM & {\sc marcs} \\
\noalign{\smallskip}
\colrule
\noalign{\smallskip}
$$[C\,{\sc i}] 		&  $8.41 $~~~~~~~~~       & $8.40 $~~~~~~~~~         & $8.41 $~~~~~~~~~         & $8.38 $~~~~~~~~~ \\
C\,{\sc i}   			& $8.42 \pm 0.05$ & $8.47 \pm 0.04$ & $8.45 \pm 0.04$ & $8.39 \pm 0.04$ \\
CH $\Delta v = 1$   	& $8.44 \pm 0.04$ & $8.44 \pm 0.04$ & $8.53 \pm 0.04$ & $8.44 \pm 0.04$ \\
CH A-X       		& $8.43 \pm 0.03$ & $8.42 \pm 0.03$ & $8.51 \pm 0.03$ & $8.40 \pm 0.03$ \\
C$_2$ Swan   		& $8.46 \pm 0.03$ & $8.46 \pm 0.03$ & $8.51 \pm 0.03$ & $8.46 \pm 0.03$ \\
\noalign{\smallskip}
\colrule
\noalign{\smallskip}
N\,{\sc i}   			& $7.78 \pm 0.04$ & $7.89 \pm 0.04$ & $7.88 \pm 0.04$ & $7.78 \pm 0.04$ \\
NH $\Delta v = 0$   	& $7.83 \pm 0.03$ & $7.94 \pm 0.02$ & $8.02 \pm 0.02$ & $7.97 \pm 0.02$ \\
NH $\Delta v = 1$   	& $7.88 \pm 0.03$ & $7.91 \pm 0.03$ & $8.01 \pm 0.03$ & $7.91 \pm 0.03$ \\
\noalign{\smallskip}
\colrule
\noalign{\smallskip}
[O\,{\sc i}] 			& $8.70 \pm 0.05$ & $8.70 \pm 0.05$ & $8.73 \pm 0.05$ & $8.69 \pm 0.05$ \\
O\,{\sc i}   			& $8.69 \pm 0.05$ & $8.73 \pm 0.05$ & $8.69 \pm 0.05$ & $8.62 \pm 0.05$ \\
OH $\Delta v = 0$   	& $8.69 \pm 0.03$ & $8.75 \pm 0.03$ & $8.83 \pm 0.03$ & $8.78 \pm 0.03$ \\
OH $\Delta v = 1$   	& $8.69 \pm 0.03$ & $8.74 \pm 0.03$ & $8.86 \pm 0.03$ & $8.75 \pm 0.03$ \\
\noalign{\smallskip}
\botrule
\end{tabular}
\end{center}
\end{table}

\newpage

\begin{table}[!t]
\caption{Representative isotopic abundance fractions in the solar system. Most of the
isotopic values are taken from 
\citet{1998JPCRD..27.1275R} 
with updates for some elements, as discussed in Sect. \ref{s:isotopes}.
\label{t:isotopes}}
\smallskip
\footnotesize
\begin{tabular}{|p{ 1mm}p{1mm}p{9.5mm}|p{ 1mm}p{1mm}p{9.5mm}|p{ 1mm}p{1mm}p{9.5mm}|p{ 1mm}p{1mm}p{9.5mm}|p{ 1mm}p{1mm}p{9.5mm}|}
\toprule
\toprule
\noalign{\smallskip}
$Z$ & $A$ & \% & $Z$ & $A$ & \% & $Z$ & $A$ & \%& $Z$ & $A$ & \% & $Z$ & $A$ & \% \\
\toprule
\noalign{\smallskip}
\noalign{\smallskip}
H  & 1   & $99.998$  & S  & 32 & $94.93$   & Fe & 57 & $2.119$   & Kr & 82  & $11.655$ & Pd & 105 & $22.33$\\
   & 2   & $ 0.002$  &    & 33 & $0.76$    &    & 58 & $0.282$   &    & 83  & $11.546$ &    & 106 & $27.33$\\
   &     &           &    & 34 & $4.29$    &    &    &           &    & 84  & $56.903$ &    & 108 & $26.46$\\
He & 3   & $0.0166$  &    & 36 & $0.02$    & Co & 59 & $100.0$   &    & 86  & $17.208$ &    & 110 & $11.72$\\
   & 4   & $99.9834$ &    &    &           &    &    &           &    &     &          &    &     &        \\
   &     &           & Cl & 35 & $75.78$   & Ni & 58 & $68.0769$ & Rb & 85  & $70.844$ & Ag & 107 & $51.839$\\
Li & 6   & $7.59$    &    & 37 & $24.22$   &    & 60 & $26.2231$ &    & 87  & $29.156$ &    & 109 & $48.161$\\
   & 7   & $92.41$   &    &    &           &    & 61 & $1.1399$  &    &     &          &    &     &        \\
   &     &           & Ar & 36 & $84.5946$ &    & 62 & $3.6345$  & Sr & 84  & $0.5580$ & Cd & 106 & $1.25$\\
Be & 9   & $100.0$   &    & 38 & $15.3808$ &    & 64 & $0.9256$  &    & 86  & $9.8678$ &    & 108 & $0.89$\\
   &     &           &    & 40 & $0.0246$  &    &    &           &    & 87  & $6.8961$ &    & 110 & $12.49$\\
B  & 10  & $19.9$    &    &    &           & Cu & 63 & $69.17$   &    & 88  & $82.6781$&    & 111 & $12.80$\\
   & 11  & $80.1$    & K  & 39 & $93.132$  &    & 65 & $30.83$   &    &     &          &    & 112 & $24.13$\\
   &     &           &    & 40 & $0.147$   &    &    &           & Y  & 89  & $100.0$  &    & 113 & $12.22$\\
C  & 12  & $98.8938$ &    & 41 & $6.721$   & Zn & 64 & $48.63$   &    &     &          &    & 114 & $28.73$\\
   & 13  & $1.1062$  &    &    &           &    & 66 & $27.90$   & Zr & 90  & $51.45$  &    & 116 & $7.49$\\
   &     &           & Ca & 40 & $96.941$  &    & 67 & $4.10$    &    & 91  & $11.22$  &    &     &        \\
N  & 14  & $99.771$  &    & 42 & $0.647$   &    & 68 & $18.75$   &    & 92  & $17.15$  & In & 113 & $4.29$\\
   & 15  & $0.229$   &    & 43 & $0.135$   &    & 70 & $0.62$    &    & 94  & $17.38$  &    & 115 & $95.71$\\
   &     &           &    & 44 & $2.086$   &    &    &           &    & 96  & $2.80$   &    &     &        \\
O  & 16  & $99.7621$ &    & 46 & $0.004$   & Ga & 69 & $60.108$  &    &     &          & Sn & 112 & $0.97$\\
   & 17  & $0.0379$  &    & 48 & $0.187$   &    & 71 & $39.892$  & Nb & 93  & $100.0$  &    & 114 & $0.66$\\
   & 18  & $0.2000$  &    &    &           &    &    &           &    &     &          &    & 115 & $0.34$\\
   &     &           & Sc & 45 & $100.0$   & Ge & 70 & $20.84$   & Mo & 92  & $14.525$ &    & 116 & $14.54$\\
F  & 19  & $100.0$   &    &    &           &    & 72 & $27.54$   &    & 94  & $9.151$  &    & 117 & $7.68$\\
   &     &           & Ti & 46 & $8.25$    &    & 73 & $7.73$    &    & 95  & $15.838$ &    & 118 & $24.22$\\
Ne & 20  & $92.9431$ &    & 47 & $7.44$    &    & 74 & $36.28$   &    & 96  & $16.672$ &    & 119 & $8.59$\\
   & 21  & $0.2228$  &    & 48 & $73.72$   &    & 76 & $7.61$    &    & 97  & $9.599$  &    & 120 & $32.58$\\
   & 22  & $6.8341$  &    & 49 & $5.41$    &    &    &           &    & 98  & $24.391$ &    & 122 & $4.63$\\
   &     &           &    & 50 & $5.18$    & As & 75 & $100.0$   &    & 100 & $9.824$   &    & 124 & $5.79$\\
Na & 23  & $100.0$   &    &    &           &    &    &           &    &     &          &    &     &        \\
   &     &           & V  & 50 & $0.250$   & Se & 74 & $0.89$    & Ru & 96  & $5.54$   & Sb & 121 & $57.21$\\
Mg & 24  & $78.99$   &    & 51 & $99.750$  &    & 76 & $9.37$    &    & 98  & $1.87$   &    & 123 & $42.79$\\
   & 25  & $10.00$   &    &    &           &    & 77 & $7.63$    &    & 99  & $12.76$  &    &     &        \\
   & 26  & $11.01$   & Cr & 50 & $4.345$   &    & 78 & $23.77$   &    & 100 & $12.60$  & Te & 120 & $0.09$\\
   &     &           &    & 52 & $83.789$  &    & 80 & $49.61$   &    & 101 & $17.06$  &    & 122 & $2.55$\\
Al & 27  & $100.0$   &    & 53 & $9.501$   &    & 82 & $8.73$    &    & 102 & $31.55$  &    & 123 & $0.89$\\
   &     &           &    & 54 & $2.365$   &    &    &           &    & 104 & $18.62$  &    & 124 & $4.74$\\
Si & 28  & $92.2297$ &    &    &           & Br & 79 & $50.69$   &    &     &          &    & 125 & $7.07$\\
   & 29  & $4.6832$  & Mn & 55 & $100.0$   &    & 81 & $49.31$   & Rh & 103 & $100.0$  &    & 126 & $18.84$\\
   & 30  & $3.0872$  &    &   &            &    &    &           &    &     &          &    & 128 & $31.74$\\
   &     &           & Fe & 54 & $5.845$   & Kr & 78 & $0.362$   & Pd & 102 & $1.02$   &    & 130 & $34.08$\\
P  & 31  & $100.0$   &    & 56 & $91.754$  &    & 80 & $2.326$   &    & 104 & $11.14$  &    &     &        \\
\noalign{\smallskip}
\botrule
\end{tabular}
\end{table}

\newpage

\begin{table}[!t]
\smallskip
\footnotesize
\begin{tabular}{|p{1mm}p{1mm}p{9.5mm}|p{1mm}p{1mm}p{9.5mm}|p{ 1mm}p{1mm}p{9.5mm}|p{ 1mm}p{1mm}p{9.5mm}|p{ 1mm}p{1mm}p{9.5mm}|}
\toprule
\toprule
\noalign{\smallskip}
$Z$ & $A$ & \% & $Z$ & $A$ & \% & $Z$ & $A$ & \%& $Z$ & $A$ & \% & $Z$ & $A$ & \% \\
\toprule
\noalign{\smallskip}
\noalign{\smallskip}
I  & 127 & $100.0$  & Nd & 142 & $27.044$& Dy & 160 & $2.329$   & Hf & 178 & $27.297$ & Pt & 196 & $25.242$ \\
   &     &          &    & 143 & $12.023$&    & 161 & $18.889$  &    & 179 & $13.629$ &    & 198 & $7.163$  \\
Xe & 124 & $0.122$  &    & 144 & $23.729$&    & 162 & $25.475$  &    & 180 & $35.100$ &    &     &          \\
   & 126 & $0.108$  &    & 145 & $8.763$ &    & 163 & $24.896$  &    &     &          & Au & 197 & $100.0$  \\
   & 128 & $2.188$  &    & 146 & $17.130$&    & 164 & $28.260$  & Ta & 180 & $0.012$  &    &     &          \\
   & 129 & $27.255$ &    & 148 & $5.716$ &    &     &           &    & 181 & $99.988$ & Hg & 196 & $0.15$   \\
   & 130 & $4.376$  &    & 150 & $5.596$ & Ho & 165 & $100.0$   &    &     &          &    & 198 & $9.97$   \\
   & 131 & $21.693$ &    &     &         &    &     &           & W  & 180 & $0.12$   &    & 199 & $16.87$  \\
   & 132 & $26.514$ & Sm & 144 & $3.07$  & Er & 162 & $0.139$   &    & 182 & $26.50$  &    & 200 & $23.10$  \\
   & 134 & $9.790$  &    & 147 & $14.99$ &    & 164 & $1.601$   &    & 183 & $14.31$  &    & 201 & $13.18$  \\
   & 136 & $7.954$  &    & 148 & $11.24$ &    & 166 & $33.503$  &    & 184 & $30.64$  &    & 202 & $29.86$  \\
   &     &          &    & 149 & $13.82$ &    & 167 & $22.869$  &    & 186 & $28.43$  &    & 204 & $6.87$   \\
Cs & 133 & $100.0$  &    & 150 & $7.38$  &    & 168 & $26.978$  &    &     &          &    &     &          \\
   &     &          &    & 152 & $26.75$ &    & 170 & $14.910$  & Re & 185 & $35.662$ & Tl & 203 & $29.524$ \\
Ba & 130 & $0.106$  &    & 154 & $22.75$ &    &     &           &    & 187 & $64.338$ &    & 205 & $70.476$ \\
   & 132 & $0.101$  &    &     &         & Tm & 169 & $100.0$   &    &     &          &    &     &          \\
   & 134 & $2.417$  & Eu & 151 & $47.81$ &    &     &           & Os & 184 & $0.020$  & Pb & 204 & $1.997$  \\
   & 135 & $6.592$  &    & 153 & $52.19$ & Yb & 168 & $0.12$    &    & 186 & $1.598$  &    & 206 & $18.582$ \\
   & 136 & $7.854$  &    &     &         &    & 170 & $2.98$    &    & 187 & $1.271$  &    & 207 & $20.563$ \\
   & 137 & $11.232$ & Gd & 152 & $0.20$  &    & 171 & $14.09$   &    & 188 & $13.337$ &    & 208 & $58.858$ \\
   & 138 & $71.698$ &    & 154 & $2.18$  &    & 172 & $21.69$   &    & 189 & $16.261$ &    &     &          \\
   &     &          &    & 155 & $14.80$ &    & 173 & $16.10$   &    & 190 & $26.444$ & Bi & 209 & $100.0$  \\
La & 138 & $0.091$  &    & 156 & $20.47$ &    & 174 & $32.03$   &    & 192 & $41.070$ &    &     &          \\
   & 139 & $99.909$ &    & 157 & $15.65$ &    & 176 & $13.00$   &    &     &          & Th & 232 & $100.0$  \\
   &     &          &    & 158 & $24.84$ &    &     &           & Ir & 191 & $37.3$   &    &     &          \\
Ce & 136 & $0.185$  &    & 160 & $21.86$ & Lu & 175 & $97.1795$ &    & 193 & $62.7$   & U  & 234 & $0.002$  \\
   & 138 & $0.251$  &    &     &         &    & 176 & $2.8205$  &    &     &          &    & 235 & $24.286$ \\
   & 140 & $88.450$ & Tb & 159 & $100.0$ &    &     &           & Pt & 190 & $0.014$  &    & 238 & $75.712$ \\
   & 142 & $11.114$ &    &     &         & Hf & 174 & $0.162$   &    & 192 & $0.782$  &    &     &          \\
   &     &          & Dy & 156 & $0.056$ &    & 176 & $5.206$   &    & 194 & $32.967$ &    &     &          \\
Pr & 141 & $100.0$  &    & 158 & $0.095$ &    & 177 & $18.606$  &
 & 195  & $33.832$ &       &      &    \\
\noalign{\smallskip}
\botrule
\end{tabular}
\end{table}

\newpage

\begin{table}[!t]
\caption{The mass fractions of hydrogen (X), helium (Y) and metals (Z) for a number of
widely-used compilations of the solar chemical composition. 
\label{t:metallicity}}
\smallskip
\begin{tabular}{lcccc}
\toprule
\toprule
\noalign{\smallskip}
Source & $X$ & $Y$ & $Z$ & $Z/X$ \\
\toprule
\noalign{\smallskip}
\noalign{\smallskip}
{\bf Present-day photosphere: } \\
\citet{1989GeCoA..53..197A}$^{\rm a}$ 	&  0.7314 	& 0.2485  	& 0.0201	& 0.0274 \\ 
\citet{1993oee..conf...15G}$^{\rm a}$ 	& 0.7336  	& 0.2485 	& 0.0179	& 0.0244 \\
\citet{1998SSRv...85..161G} 	& 0.7345	& 0.2485 	& 0.0169 	& 0.0231 \\
\citet{2003ApJ...591.1220L} 	& 0.7491	& 0.2377 	& 0.0133 	& 0.0177 \\  
\citet{2005ASPC..336...25A} 	& 0.7392	& 0.2485	& 0.0122	& 0.0165 \\
\citet{2009arXiv0901.1149L} 			& 0.7390	& 0.2469 	& 0.0141 	& 0.0191 \\  
Present work				& 0.7381	& 0.2485	& 0.0134	& 0.0181\\
\noalign{\smallskip}
{\bf Proto-solar: } \\
\citet{1989GeCoA..53..197A} 	& 0.7096 	& 0.2691	& 0.0213	& 0.0301 \\
\citet{1993oee..conf...15G} 	& 0.7112	& 0.2697	& 0.0190 	& 0.0268 \\
\citet{1998SSRv...85..161G} 	& 0.7120	& 0.2701	& 0.0180	& 0.0253 \\
\citet{2003ApJ...591.1220L} 	& 0.7111	& 0.2741	& 0.0149	& 0.0210 \\ 
\citet{2005ASPC..336...25A} 	& 0.7166	& 0.2704	& 0.0130	& 0.0181 \\
\citet{2009arXiv0901.1149L} 			& 0.7112	& 0.2735	& 0.0153	& 0.0215 \\ 
Present work				& 0.7154	& 0.2703	& 0.0142	& 0.0199	\\
\noalign{\smallskip}
\botrule
\end{tabular}
\\
$^{\rm a}$ The He abundances given in \citet{1989GeCoA..53..197A} and
\citet{1993oee..conf...15G} 
have here been replaced with the current
best estimate from helioseismology (Sect. \ref{s:noble}).
\end{table}

\newpage

\begin{table}[!t]
\caption{Comparison of the proto-solar abundances from the present work and
\citet{1998SSRv...85..161G} 
with those in nearby B stars and H\,{\sc ii} regions. 
The solar values given here include the effects of diffusion
\citep{2002JGRA..107.1442T} 
as discussed in Sect. \ref{s:bulk}.
The H\,{\sc ii} numbers include the estimated elemental fractions tied up in dust; 
the dust corrections for Mg, Si and Fe are very large and thus too uncertain 
to provide meaningful values here.  
Also given in the last column is the predicted Galactic chemical enrichment (GCE)
over the past 4.56\,Gyr.
\label{t:neighborhood}}
\smallskip
\begin{tabular}{lccccc}
\toprule
\toprule
\noalign{\smallskip}
Elem. &  Sun$^{\rm a}$  & Sun$^{\rm b}$  & B stars$^{\rm c}$ & H\,{\sc ii}$^{\rm d}$  & GCE$^{\rm e}$  \\
\toprule
\noalign{\smallskip}
\noalign{\smallskip}
He & $10.98\pm0.01$ & $10.98\pm0.01$ &  $10.98\pm0.02$ & $10.96\pm0.01$ & 0.01 \\
C   & $8.56\pm0.06$   & $8.47\pm0.05$    &  $8.35\pm0.03$   & $8.66\pm0.06$   & 0.06 \\
N   & $7.96\pm0.06$   & $7.87\pm0.05$    &  $7.76\pm0.05$   & $7.85\pm0.06$   & 0.08 \\
O   & $8.87\pm0.06$   & $8.73\pm0.05$    &  $8.76\pm0.03$   & $8.80\pm0.04$   & 0.04 \\
Ne & $8.12\pm0.06$   & $7.97\pm0.10$    &  $8.08\pm0.03$   & $8.00\pm0.08$   & 0.04 \\
Mg & $7.62\pm0.05$   & $7.64\pm0.04$   &  $7.56\pm0.05$   &                                & 0.04 \\
Si   & $7.59\pm0.05$   & $7.55\pm0.04$   &  $7.50\pm0.02$   &                                & 0.08 \\
S    & $7.37\pm0.11$   & $7.16\pm0.03$   &  $7.21\pm0.13$   &  $7.30\pm0.04$  & 0.09 \\
Ar   & $6.44\pm0.06$   & $6.44\pm0.13$   &  $6.66\pm0.06$   &  $6.62\pm0.06$  &  \\
Fe  & $7.55\pm0.05$   & $7.54\pm0.04$   &  $7.44\pm0.04$   &                                & 0.14 \\
\noalign{\smallskip}
\botrule
\end{tabular}
\\
$^{\rm a}$ \citet{1998SSRv...85..161G} 
$^{\rm b}$ Present work
$^{\rm c}$ \citet{2008ApJ...688L.103P}, 
\citet{2006A&A...457..651M}, 
\citet{2008ApJ...678.1342L} 
$^{\rm d}$ \citet{2004MNRAS.355..229E, 2005ApJ...618L..95E}, 
\citet{ 2007ApJ...670..457G} 
$^{\rm e}$ \citet{2003MNRAS.339...63C}. 
\end{table}

\newpage

\clearpage

\clearpage

\clearpage

\clearpage

\clearpage

\clearpage

\clearpage

\end{document}